\documentclass[fleqn,usenatbib]{mnras}

\usepackage{newtxtext,newtxmath}
\usepackage[T1]{fontenc}

\DeclareRobustCommand{\VAN}[3]{#2}
\let\VANthebibliography\thebibliography
\def\thebibliography{\DeclareRobustCommand{\VAN}[3]{##3}\VANthebibliography}

\usepackage{graphicx}	% Including figure files
\usepackage{amsmath}	% Advanced maths commands
\title[Early metal enrichment]
{The extent of intergalactic metal enrichment from galactic winds during the Cosmic Dawn}

\author[Yamaguchi, Furlanetto, \& Trapp]{
Natsuko Yamaguchi,$^{1,2}$\thanks{E-mail:nyamaguc@caltech.edu}
Steven R. Furlanetto$^{1}$
, \& A.~C. Trapp,$^1$ 
\\
$^{1}$Department of Physics and Astronomy, University of California Los Angeles, CA, 90095-1562, USA \\
$^{2}$Cahill Center for Astronomy and Astrophysics, California Institute of Technology, Pasadena CA 91125, USA
}

\date{Accepted XXX. Received YYY; in original form ZZZ}

\pubyear{2021}

\begin{document}
\label{firstpage}
\pagerange{\pageref{firstpage}--\pageref{lastpage}}
\maketitle

% Abstract of the paper
\begin{abstract}

One of the key processes driving galaxy evolution during the Cosmic Dawn is supernova feedback. This likely helps regulate star formation inside of galaxies, but it can also drive winds that influence the large-scale intergalactic medium.
Here, we present a simple semi-analytic model of supernova-driven galactic winds and explore the contributions of different phases of galaxy evolution to cosmic metal enrichment in the high-redshift $(z \gtrsim 6)$ Universe. We show that models calibrated to the observed galaxy luminosity function at $z \sim 6$--8 have filling factors $\sim 1\%$ at $z \sim 6$ and $\sim 0.1\%$ at $z \sim 12$, with different star formation prescriptions providing about an order of magnitude uncertainty. Despite the small fraction of space filled by winds, these scenarios predict an upper limit to the abundance of metal-line absorbers in quasar spectra at $z \ga 5$ which is comfortably above that currently observed. We also consider enrichment through winds driven by Pop III star formation in minihalos. We find that these  can dominate the total filling factor at $z \ga 10$ and even compete with winds from normal galaxies at $z \sim 6$, at least in terms of the total enriched volume. But these regions have much lower overall metallicities, because each one is generated by a small burst of star formation. Finally, we show that Compton cooling of these supernova-driven winds at $z \ga 6$ has only a small effect on the cosmic microwave background. 

\noindent 
\end{abstract}

% Select between one and six entries from the list of approved keywords.
% Don't make up new ones.
\begin{keywords}
galaxies: high-redshift -- intergalactic medium -- dark ages, reionization, first stars
\end{keywords}

%%%%%%%%%%%%%%%%%%%%%%%%%%%%%%%%%%%%%%%%%%%%%%%%%%

%%%%%%%%%%%%%%%%% BODY OF PAPER %%%%%%%%%%%%%%%%%%

%%%%%%%%%%%%%%%%%%%%%%%%%%%%%%%%%%%%%%%%%%%%%%%%%%%%%%%%%%%%%%%%
%%%%%%%%%%%%%%%%%%%%%%%%%%%%%%%%%%%%%%%%%%%%%%%%%%%%%%%%%%%%%%%%

\section{Introduction}

High-redshift galaxy formation and evolution at $z \geq 6$ have been amongst the primary foci of extragalactic astrophysics for the past two decades. In particular, the Cosmic Dawn era, beginning just a few hundred million years after the Big Bang, is a crucial time period that saw the birth of the first stars. The Cosmic Dawn not only ended the Dark Ages, but these early populations of stars were the progenitors of all the structure that we see today and would later become the dominant drivers of reionization, the last global phase transition of the Universe.

Despite the challenge of probing such a distant Universe, in recent years there has been substantial improvement in our census of luminous objects at $z \sim 6$--$8$ \citep{mclure_new_2013, schenker_uv_2013, schmidt_luminosity_2014, bouwens_uv_2015, steven_l_finkelstein_evolution_2015, atek_are_2015, bowler_unveiling_2017, livermore_directly_2017}, which has allowed us to develop a good understanding of bright galaxies at these times. 
Higher redshifts are just now beginning to be explored by JWST, but the situation is not yet clear \citep{oesch_remarkably_2016, donnan_evolution_2022, castellano_early_2022,  harikane_comprehensive_2022}. 

Nonetheless, there have been numerous recent efforts to model various aspects of galaxy evolution in this era (\citealt{furlanetto_minimalist_2017}; \citealt{furlanetto_bursty_2021}; \citealt{moster_emerge_2018}; \citealt{popping_dust_2017}; also see \citet{dayal_early_2018} for an overview of the physics of early galaxy formation and the types of theoretical tools used to model it). In general, these models show that the properties of observed galaxies at $6 \lesssim z \lesssim 8$ can be understood in a similar fashion to galaxies at later times. For example, models in which stellar feedback regulates star formation can quite successfully explain the observed luminosity function \citep{mason_galaxy_2015, furlanetto_minimalist_2017, mirocha_global_2017}. 

One particularly critical aspect of this feedback is the presence of ``winds" driven by supernovae (SNe). Cosmological hydrodynamical simulations incorporating SN feedback to study galaxy properties and populations include, e.g., \citet{scannapieco_highredshift_2001, springel_cosmological_2003, crain_eagle_2015, fierlinger_stellar_2016, dave_simba_2019}, while those concerned primarily with its contribution to metallicity and/or metal enrichment include, e.g., \citet{theuns_galactic_2002, kobayashi_simulations_2007, finlator_origin_2008, suresh_impact_2015, nelson_first_2019}. Meanwhile, \citet{matteucci_relative_1986} and \citet{white_galaxy_1991} were amongst the first to introduce semi-analytic approaches to study chemical enrichment. Later examples modeling z $\lesssim$ 3 galaxies include \citet{somerville_semi-analytic_1999, henriques_simulations_2013, hayward_how_2017} while those extending to z $\gtrsim$ 3 include \citet{nath_enrichment_1997, ferrara_mixing_2000, madau_early_2001, de_lucia_chemical_2004}. Many of these studies show how such supernova explosions can help regulate star formation inside galaxies. But beyond the host galaxy, these winds can also influence the surrounding intergalactic medium (IGM) by ejecting material into it. A significant consequence of this in the early Universe is metal enrichment over large volumes of the Universe -- including gas that will later be accreted onto growing galaxies. Though there are many other processes that may have contributed to widespread enrichment (e.g. quasar winds, radiation pressure-driven dust outflows), winds from star-forming galaxies are the clearest culprit for polluting the large volumes of the Universe that we see today \citep{ferrara_mixing_2000,madau_early_2001, furlanetto_metal_2003}. 

Existing observations have shown that metals are relatively widespread by $z \sim 3$. For example, optical spectra of Lyman-break galaxies  indicate substantial enrichment \citet{shapley_rest-frame_2003, adelberger_galaxies_2003}, as do radio measurements  \citet{ginolfi_alpine-alma_2020}, while enrichment inside galaxy clusters is also well-known \citet{tozzi_iron_2003, baldi_xmm-newton_2012, mcdonald_evolution_2016}. Meanwhile, Ly-$\alpha$ forest and related techniques require even more widespread metals \citet{songaila_metal_1996, cowie_heavy-element_1998, ellison_enrichment_2000, schaye_detection_2000, crane_ionization_1995}, at least outside of underdense voids in the galaxy distribution (where the observations cannot yet probe). More recent observations revealed metal absorbers existing by $z \sim 5-6$. For example,  \citet{ryan-weber_intergalactic_2006} and \citet{becker_discovery_2006} reported absorption of [CIV] in $z \sim 5$ and [OI] in $z \sim 6$ QSO systems respectively, while  \citet{sparre_metallicity_2014} discovered Fe II fine line structures in the afterglow of GRBs at $z \sim 5$. However, only a handful of absorbers are known above $z \ga 5$, and it is difficult to determine how widespread the metal enrichment was during this era. 

Meanwhile, it is clear that high-$z$ galaxies are themselves significantly enriched: \citet{capak_galaxies_2015} detected [CII] gas emission in star forming galaxies at $z \sim 5-6$, and  \citet{faisst_rest-uv_2016} found that the average galaxy population at $z \sim 5$ has a metallicity comparable to those at $z \sim 3.5$. Therefore, it is becoming increasingly evident that metal enrichment started in the very early stages of the bright Universe, though its precise timeline remains unclear. A crucial uncertainty lies in the phase of galaxy evolution at which metals are ejected -- whether it begins with the birth and death of Population III stars or as part of ``normal" feedback regulation at later times. 

Interpretation of the handful of metal-line measurements presents an important challenge. While the overall metal production rate depends mostly on the star formation rate (but also on stellar models), the spatial extent of metal enrichment
%, which is often described through the volume filling factor (or the fraction of the universe enriched by metals at a given time), 
is necessary to understand the impact of these metals. There have been many attempts to estimate this factor in the past, but early calculations were not based on realistic galaxy populations, which likely resulted in an overestimate of the filling factor (or the fraction of the Universe enriched by metals at a given time; e.g.,  \citealt{madau_early_2001, furlanetto_metal_2003}). More recent calculations have typically used numerical simulations, so it is difficult to assess the uncertainties associated with them (e.g., \citealt{jaacks_baseline_2018}).  The recent launch of JWST, which will make extraordinary progress in understanding metal lines during this era \citep{mason_galaxy_2015, yung_semi-analytic_2019-1, vogelsberger_high-redshift_2020}, makes modeling the process very timely. 

The extent of metal enrichment is important both for understanding galaxy evolution (as the loss of metals affects the chemical evolution of the stars and interstellar medium)
%(as the IGM gas provides the fuel for later star formation) 
and also for probes of reionization and the radiation fields at early times. For galaxy evolution, the extent of the winds can be used to constrain models of stellar feedback. The  evolution of the radiation background determines the ionization state of the metals, so in principle quasar metal lines are powerful probes of reionization \citep{oh_probing_2002, hennawi_probing_2021}, but only if the distribution of metals themselves is known as well (or at least can be jointly measured). 
In this paper, we address this question by pairing a simple galaxy evolution model with a simple wind model. In particular, we use the minimalist model for feedback regulation described in \citet{furlanetto_minimalist_2017} to calibrate the galaxy parameters to reproduce observed luminosity functions.  We then follow the resulting winds using the model of \citet{furlanetto_metal_2003} (which is in turn based upon \citealt{tegmark_late_1993}). Although such a simplistic model cannot follow star formation or wind propagation in as detailed a manner as numerical simulations, its flexibility allows for a broader study of the many uncertainties in early galaxy formation. We explore how the overall volume filling factor and absorption line statistics at $z \gtrsim 4$ depend upon assumptions about star-forming galaxies during this era. As it remains difficult to detect metals at $z \geq 6$, we also consider distortions to the cosmic microwave background (CMB) through the Sunyaev–Zel'dovich effect as an alternative observable for such winds \citep{oh_sunyaev-zeldovich_2003}.

This paper is organized as follows. In Section \ref{sec:galaxy_model}, we introduce our galaxy evolution model, including a description of the underlying dark matter halo population and our prescription for feedback regulation. In Section \ref{sec:wind_model}, we present our wind model and show its solutions for individual galactic winds. We then discuss the implications of our model for the extent of metal enrichment, as well as potential probes of it, in Section \ref{sec:results}. Finally, we conclude in Section \ref{sec:discussion}.

In accordance with the Planck Collaboration XIII results \citep{planck_collaboration_planck_2016}, we use the following cosmological parameters: $\Omega_{m} = 0.308$, $\Omega_{\Lambda} = 0.692$, $\Omega_{b} = 0.0484$, and $h = 0.678$. For astrophysical constants, we use those found in Section~2 of \citet{beringer_review_2012}.

\section{A model for star-forming galaxies} \label{sec:galaxy_model}

In this section, we introduce our simple galaxy model. We refer the reader to \citet{furlanetto_minimalist_2017} for more details.

\subsection{Dark matter halos} \label{ssec: darkmatterhalos}

We define the co-moving number density of dark matter halos in the mass range
$\left(M_h, M_h+\textup{d}M_h\right)$ at redshift $z$ as $n_h\left(M_h,z\right)\textup{d} M_h$. By convention, we write
\begin{equation}
    n_h\left(M_h,z\right) = f\left(\sigma\right)\frac{\bar{\rho}}{M_h}\frac{d \ln\left(1/\sigma\right)}{\textup{d}M_h}
\end{equation}
where $\bar{\rho}$ is the comoving average matter density, $\sigma\left(M, z\right)$ is the linear rms fluctuation of the matter density field smoothed on a scale $M$, 
and $f\left(\sigma\right)$ is a dimensionless function taken from a fit to high-$z$ cosmological simulations \citep{trac_scorch_2015}:
\begin{equation}
    f\left(\sigma\right) = 0.150\left[1+\left(\frac{\sigma}{2.54}\right)^{-1.36}\right]e^{-{1.14}/{\sigma^2}}.
\end{equation}
We note that this mass function has not been verified at the highest redshifts and smallest masses relevant to our Pop III model (see section~\ref{ssec: burst}), but \citet{mebane_persistence_2018} found that the differences with other mass functions were modest compared to the overall uncertainties in star formation during this era.

Our galaxy model requires the accretion rate onto dark matter haloes. We use the method described in \citet{furlanetto_minimalist_2017}, which assumes that accretion proceeds smoothly without mergers (this is appropriate as our models are primarily concerned with the total amount of star formation over the history of the halos and thus not very sensitive to instantaneous scatter in accretion rate). In analogy with abundance matching \citep{vale_linking_2004}, we demand that the overall number density of the halos remains constant, with each halo growing in order to maintain the underlying mass function.  We assume that the halos remain in the same relative ordering in mass as well. In other words,  at any two redshifts $z_1$ and $z_2$,
\begin{equation}
    \int_{M_1}^{\infty}\textup{d}M\ n_h\left(M\vert z_1\right) = \int_{M_2}^{\infty}\textup{d}M\ n_h\left(M\vert z_2\right),
\end{equation}
where $M_1$ and $M_2$ are the respective masses of a given halo at the two redshifts. The accretion rate, $\dot{M}_h$, is then obtained by requiring that this relation be satisfied at all redshifts and all halo masses. 

We must also specify the range of halo masses allowed to form stars (and hence drive winds).  This threshold is determined by two conditions: (1) the halo must exceed the ``filter mass" \citep{gnedin_probing_1998} above which baryons can accrete and (2) the halo gas must cool efficiently after virialization. Motivated by the second condition, for most of our models we take this minimum mass to correspond to a virial temperature of $10^4$~K, above which atomic cooling becomes efficient. We consider winds from halos below this mass threshold in Section~\ref{ssec: burst}.

\subsection{Star formation and feedback}

Next, we describe star formation and feedback in the minimalist galaxy formation model (referred to as the ``normal" model hereafter) as well as a ``bursty" model for lower mass halos. 

The fundamental idea of this model is the assumption that stellar feedback controls the star formation rate (SFR) of each galaxy. First, we balance the rate at which baryons are accreted onto the halo, $\dot{m}_b$, with the SFR, $\dot{m}_{\star}$, and the rate of at which baryons are expelled through processes such as radiation pressure and supernovae, $\dot{m}_w$:
\begin{equation}
    \dot{m}_b = \dot{m}_{\star} + \dot{m}_w.
\end{equation}
Because the feedback is driven by star formation, we then assume $\dot{m}_w = \eta\dot{m}_{\star}$. The constant of proportionality, $\eta$, (commonly known as the mass-loading factor) is in general a function of halo mass and redshift and represents the strength of stellar feedback. 

The star formation efficiency (SFE) is defined as the fraction of the accreted baryons which form stars, $f_{\star} = \dot{m}_{\star}/\dot{m}_b$. Thus,
\begin{equation}
\label{eqn:fstar}
    f_{\star} = \frac{1}{1+\eta(M_h,z)}.
\end{equation}

However, to match the observed star formation efficiencies at large halo masses, quenching processes which suppress accretion are often invoked. Although such effects do not significantly affect our results, we will incorporate virial shock heating as one example, suppressing accretion by a factor $f_{\rm shock}$ following \citet{faucher-giguere_baryonic_2011}. We also impose an upper limit, $f_{\star,\rm max}$, on the star formation efficiency. Incorporating both these effects in such a way as to maintain continuity, we obtain
\begin{equation}
    f_{\star} = \frac{f_{\rm shock}}{f_{\star,\rm max}^{-1} + \eta(M_h,z)}. \label{fstar}
\end{equation} 

We are interested in considering a wide range of metal enrichment scenarios so include two distinct models of feedback regulation. Firstly, the ``energy-regulated'' model is derived by balancing a fixed fraction of the supernova energy (the portion in a kinetic form) with the kinetic energy of the gas that is lifted out as a wind:
\begin{equation}
    \frac{1}{2}\dot{m}_w v_{\rm esc}^2 = \dot{m}_{\star}\epsilon_k\omega_{\rm SN},  
\end{equation}
where $v_{\rm esc}$ is the halo escape velocity, $\epsilon_k$ is the fraction of the SN energy released in the wind, and $\omega_{\rm SN} \approx 10^{49}$erg/$M_{\odot}$ is the supernova energy produced per unit mass of star formation  (where the fiducial value comes from taking the typical energy released per SN, $\sim 10^{51}$ erg, and the typical number of SNe per solar mass formed, $\sim 0.015 M_{\odot}$). This results in 
\begin{equation} \label{eqn:eta}
    \eta = C \left(\frac{10^{11.5}M_{\odot}}{M_h}\right)^{\xi}\left(\frac{9}{1+z}\right)^{\sigma},
\end{equation}
with exponents $\xi = \frac{2}{3}$ and $\sigma = 1$. However, due to radiative cooling or other processes that occur in high density and temperature environments, feedback can in practice be much less efficient. Therefore, we also consider a  ``momentum-regulated'' case, balancing the momentum of the supernova blastwaves with the gas escaping the halo, done in a similar way to energy conservation above. This yields $\xi = \frac{1}{3}$ and $\sigma = \frac{1}{2}$. These scenarios provide helpful intuition for the parameters in equation~(\ref{eqn:eta}); given the difficulty of modeling the interaction of stellar feedback with the interstellar medium, we do not try to do better here (though see \citealt{hayward_how_2017, furlanetto_quasi-equilibrium_2021}). The normalization constant $C$ and the parameter $f_{\star,\rm max}$ are set by comparison to observations. 
The parameters used for each model are summarized in Table~\ref{tab:Table 1}. All of these feedback-regulated models have an SFE that increases with halo mass in faint galaxies, aside from the constant model.

For completeness, we also consider a model with a constant star formation efficiency in all halos ($\xi=0$ and $\sigma=0$), which matches the assumptions in many early treatments of metal enrichment. 

These feedback-regulated models do assume that star formation within galaxies is able to come into a quasi-equilibrium state with respect to infall. It is not obvious that this can occur at high redshifts. For example, \citet{furlanetto_bursty_2021}  showed that incorporating a time delay between star formation and supernova feedback results in oscillations (i.e. repeated ``bursts") of the SFR, particularly in lower mass halos. This moderates the dependence of the average SFE on halo mass, boosting the star formation rate of small galaxies. It has little effect on massive galaxies, because eventually feedback is no longer able to eject all of the halo gas, at which point the SFR transitions to a smooth function. To mimic this kind of out-of-equilibrium behavior, we also consider a ``bursty'' model, that imposes a minimum SFE value $f_{\star,\rm min}$ = 0.03, in which the SFE effectively decouples from the properties of the halo.
 
We emphasize that all of these parameter choices  provide reasonable fits to the observed luminosity functions (LFs) at $z \sim 6$--8, (see Fig.~3 of \citealt{furlanetto_minimalist_2017} and Fig.~8 of \citealt{furlanetto_bursty_2021}).  Importantly, the minimum SFE in the bursty model only affects faint galaxies that have not yet been observed. We emphasize that the model with constant SFE, $f_\star = 0.1$ does not produce realistic galaxy populations, at least at $z \sim 6$--8, but we include it anyway for comparison with earlier results. 
 
 \begin{table} 
     \centering
     \begin{tabular}{|c c c c c|}
          \hline \hline
          Model & $C$ & $\xi$ & $\sigma$ & $f*_{\rm max}^{-1}$ \\
          \hline
          Energy-reg & $10\epsilon_k$ & $\frac{2}{3}$ & $1$ & $0.1$ \\
          Momentum-reg & $\epsilon_p$ & $\frac{1}{3}$ & $\frac{1}{2}$ & $0.2$ \\
          \hline  
     \end{tabular}
     \caption{SFE parameters for our normal galaxy models, in order to reproduce observed luminosity functions. We also set $\epsilon_k = 0.1$ and $\epsilon_p = 5$.}
     \label{tab:Table 1}
 \end{table}

\subsection{Metal production} \label{ssec:MP}

We follow \citet{furlanetto_metal_2003} to estimate the metal production rate of our galaxies. The mass of a metal $i$ produced by a galaxy is
\begin{equation} \label{eqn:M_metal}
    M_{i} = Y_i n_{\rm SN}\frac{\Omega_b}{\Omega_m}X_\star M_h
\end{equation}
where $Y_i$ is the average yield of the element $i$ produced per Type II supernova, $n_{\rm SN} \approx 10^{-2} M_{\odot}^{-1}$ is the number of supernovae per unit mass of star formation, and $X_\star$ is the time-averaged SFE for the galaxy in question, which is slightly smaller than the instantaneous SFE $f_\star$ in our models (see \citealt{furlanetto_minimalist_2017}). 

For a comparison of enrichment to observations, we must transform the metals into a set of discrete absorption lines. In section \ref{sec:wind_model}, we will consider wind bubbles driven by supernovae in each galaxy. In this context, we then estimate the total column density of the metal along a line of sight through a wind bubble 
via 
\begin{equation}
    N_i \approx \frac{M_{i}}{m_i}\frac{1}{4\pi R^2}
\end{equation}
where $m_i = A_i u$ is the mass of a metal atom. Note that this is simply an estimate of the characteristic column density; even if the metals are distributed uniformly, the column density would vary with the impact parameter. We emphasize that this equation assumes a roughly uniform distribution of metals; if the metals were clumped, fewer lines of sight would encounter metals, but those that did would have stronger absorption. 

For an unsaturated line, the equivalent width is given by
\begin{multline} \label{eqn:EW}
    W \approx 0.8 \mbox{\AA} \, \left(\frac{f_{\rm osc}}{0.05}\frac{Y_x}{0.5 M_{\odot}}\frac{16}{A_x}\right)\left(\frac{\lambda_m}{1300\mathring{A}}\right)^2\\\left(\frac{X*}{0.1}\frac{\omega_{\rm SN}}{10^{51}\rm{ergs}/126M_{\odot}}\frac{\Omega_b/\Omega_m}{0.05/0.3}\frac{M_h}{10^9 M_{\odot}}\right)\left(\frac{0.02 \rm Mpc}{R}\right)^2\left(\frac{1+z}{10}\right)
\end{multline}
where $\lambda_{m}$ is the wavelength of the transition and $f_{\rm osc}$ is its oscillator strength. 

In Section \ref{ssec: dndx}, we compare predictions for the OI and CIV transitions to observations. We choose these because they are amongst the very few which have been observed at $z \sim 6$ and because OI is a strong line in mostly neutral gas while CIV is a strong line in mostly ionized gas. For OI, we take $Y = 0.5 \ M_{\odot}$,  $f_{\rm osc} = 0.04887$,  $\lambda_{m} = 1302.2$ \AA; while for CIV we take $Y = 0.1 \  M_{\odot}$,  $f_{\rm osc} = 0.1908$,  $\lambda_{m} = 1548.2$ \AA \space (CIV is a doublet, so these values correspond to the stronger transition, as reported in \citet{becker_high-redshift_2009} to which we compare in Section~\ref{ssec: dndx}). These average yields are those produced per Type II supernova from \citet{woosley_evolution_1995} which depends on the energy of the supernova. Following \citet{furlanetto_metal_2003}, we take the values from the lower energy model and neglect chemical evolution within the galaxies. These approximations are reasonable as uncertainties arising from them are likely less significant than those from other simplifications in our model. Note that given the short timescales involved, we do not include contributions from other mechanisms such as winds from AGB stars and Type Ia SNe. This probably results in a modest underestimate of the carbon abundance, as over $\sim 100$~Myr timescales high-mass AGB winds and even Type~Ia SNe can contribute.

\subsection{Pop~III star formation} \label{ssec: burst}

So far, we have focused on ``normal'' galaxies during the Cosmic Dawn. But many models expect another set of star-forming halos hosting Population III stars. We thus also consider potential contributions from smaller halos hosting these exotic stars.

The criteria to form Pop~III stars are complex: halos must be massive enough to accrete baryons and then form H$_2$, which is the primary coolant in this regime. But H$_2$ is fragile and is destroyed by ultraviolet photons -- which are themselves produced by Pop~III stars. Models of this era show that the minimum halo mass can therefore have a complex evolutionary history (e.g., \citealt{jaacks_baseline_2018, mebane_persistence_2018}). We therefore consider two simple models. For one, we take a minimum virial temperature of $1000$~K for this population (``constant $T_{\min}$ model''). This is a few times smaller than the ``plateau'' found in the minimum halo in \citet{mebane_persistence_2018}, so it provides an optimistic estimate of the enrichment. In addition, we also consider a prescription for $T_{\rm min}$ with a stronger time dependence inspired by an updated version of the \citet{mebane_persistence_2018} model (S.~Hegde, private communication). This includes improved estimates of halo self-shielding and dark matter--baryon streaming (as studied recently in \citealt{kulkarni_critical_2021}).  We find the following simple fit roughly matches the new model: 
\begin{equation}
    M_{\rm h0} = 0.916 e^{-0.568\left(z-38.3\right)} + 1.21 \times 10^6 \ M_\odot
\end{equation}

Such low mass halos likely have brief periods of rapid star formation that then shut off, until they are able to accrete enough mass to retain gas and begin normal star formation (e.g,  \citealt{abel_formation_2002, bromm_formation_2002}). 
Although each such halo will form only a small mass of stars, the halos are so numerous that their contribution to the overall filling factor of winds can nevertheless be substantial (especially at very early times). We model these sources as hosting a \emph{single} burst of  star formation, producing a specified total stellar mass $M_* \sim 100$--$1000 \ M_{\odot}$, after which star formation halts completely. This is in contrast to our fiducial model, in which star formation continues over long time periods. We also assume that Pop~III stars have more energetic supernovae than those in normal halos, so we take $\omega_{\rm SN} = 10^{50}$~erg/$M_\odot$, about an order of magnitude larger. 
We assume the  metal yields are the same as our fiducial model, although these could be underestimates if pair-instability supernovae dominate. Our treatment of Population~III star-forming halos is thus very approximate, but it should provide some intuition for the contribution of these kinds of halos.

\section{Wind-driven bubbles}\label{sec:wind_model}

\subsection{Wind physics}

In this section, we provide an overview of our wind expansion model, which is a simplified version of \citet{furlanetto_metal_2003} (which is itself an implementation of \citealt{tegmark_late_1993}). In the model, winds are driven by supernovae inside galaxies and commence as soon as star formation does. Once a wind forms, it expands and sweeps up the ambient IGM. While a fraction of the material forms the shell, the remaining gas enters the hot, rarefied interior, whose thermal pressure helps expand the shell. Using the thin-shell approximation and assuming spherical symmetry, the following system of equations can be used to describe the evolution of an individual wind bubble:
\begin{align}
    M_d &= M_h + \frac{4\pi}{3}\Bar{\rho}_d^0\left(1+z\right)^3R^3 \\
    \ddot{R} &= \frac{4\pi R^2}{M_s}P-\frac{G}{R^2}\left(M_d+\frac{M_s}{2}\right)-\frac{\dot{M}_s}{M_s}\left(\dot{R}-HR\right)\\
    \dot{P} &= \frac{L}{2\pi R^3}-5P\frac{\dot{R}}{R} \\
    \dot{M}_s &= \left.
    \begin{cases} 0 ,& \text{for } \dot{R} \leq HR \\ 
    4 \pi R^2 \Bar{\rho}_b^0\left(1+z\right)^3\left(\dot{R}-HR\right), & \text{for } \dot{R} > HR 
    \end{cases} 
    \right.
\end{align}
where $M_h$ is the mass of the halo in which the source galaxy resides, $M_d$ is the dark matter mass enclosed by the wind shell (with mass $M_s$), $R$ is the radius of the wind bubble, $P$ is the pressure of the bubble interior, and $L$ is the rate energy is injected to drive the wind. $\Bar{\rho}_d^0$ and $\Bar{\rho}_b^0$ are the average dark matter and baryon densities at $z = 0$, respectively. Furthermore, the accretion rate, $\dot{M}_h$, is obtained from abundance matching as described in Section \ref{ssec: darkmatterhalos}.

The luminosity term in the pressure equation is made up of two components:
\begin{equation}
    L = L_w + L_{\rm comp}.
\end{equation}
Here $L_{w}$ is the energy injected by stellar feedback and $L_{\rm comp}$ is the rate at which the thermal energy is lost to Compton cooling. For the first, 
\begin{equation}
    L_{w} = f_{\star}\frac{\Omega_b}{\Omega_m}\dot{M}_h\epsilon_k\omega_{\rm SN} \label{eqn:Lw}
\end{equation}
where $f_\star$ is determined for normal galaxies as described in Section~\ref{sec:galaxy_model}. As described in Section \ref{ssec: burst}, in the Pop III model, this component is set to zero after the initial burst of star formation.

$L_{\rm comp}$ is the energy lost through inverse Compton cooling, where energy is transferred from a hot charged particle in the wind to a cosmic microwave background (CMB) photon through scattering; while this process is very slow in the present Universe, the increased CMB energy density at $z > 6$ can make it important (and potentially observable, as we will explore in Section~\ref{ssec:CMB_effects}).  The rate energy is lost through this process is
\begin{equation}
    L_{\rm comp} = - \frac{3}{2} \frac{PV}{t_{\rm comp}},
\end{equation}
where $V$ is the volume enclosed by the wind and the cooling time is
\begin{equation}
    t_{\rm comp} = \left(\frac{8\sigma_Ta_{\rm rad}T_{\gamma}^4}{3m_e c}\right)^{-1} = 1.2 \times 10^8 \left(\frac{1+z}{10}\right)^{-4} \ {\rm yr}
\end{equation}

\begin{figure*}
    \centering
    \includegraphics[width=\columnwidth]{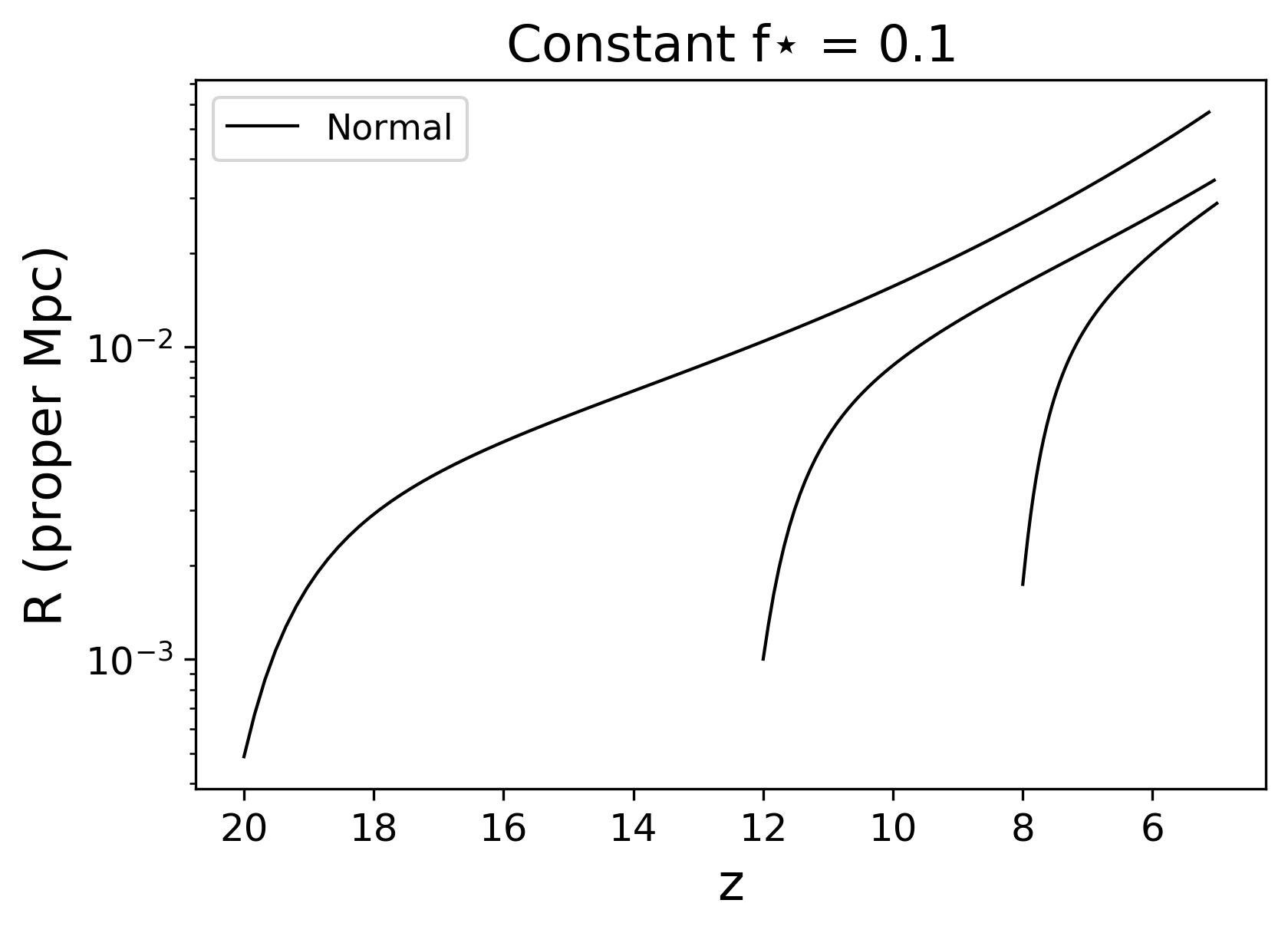}
    \includegraphics[width=\columnwidth]{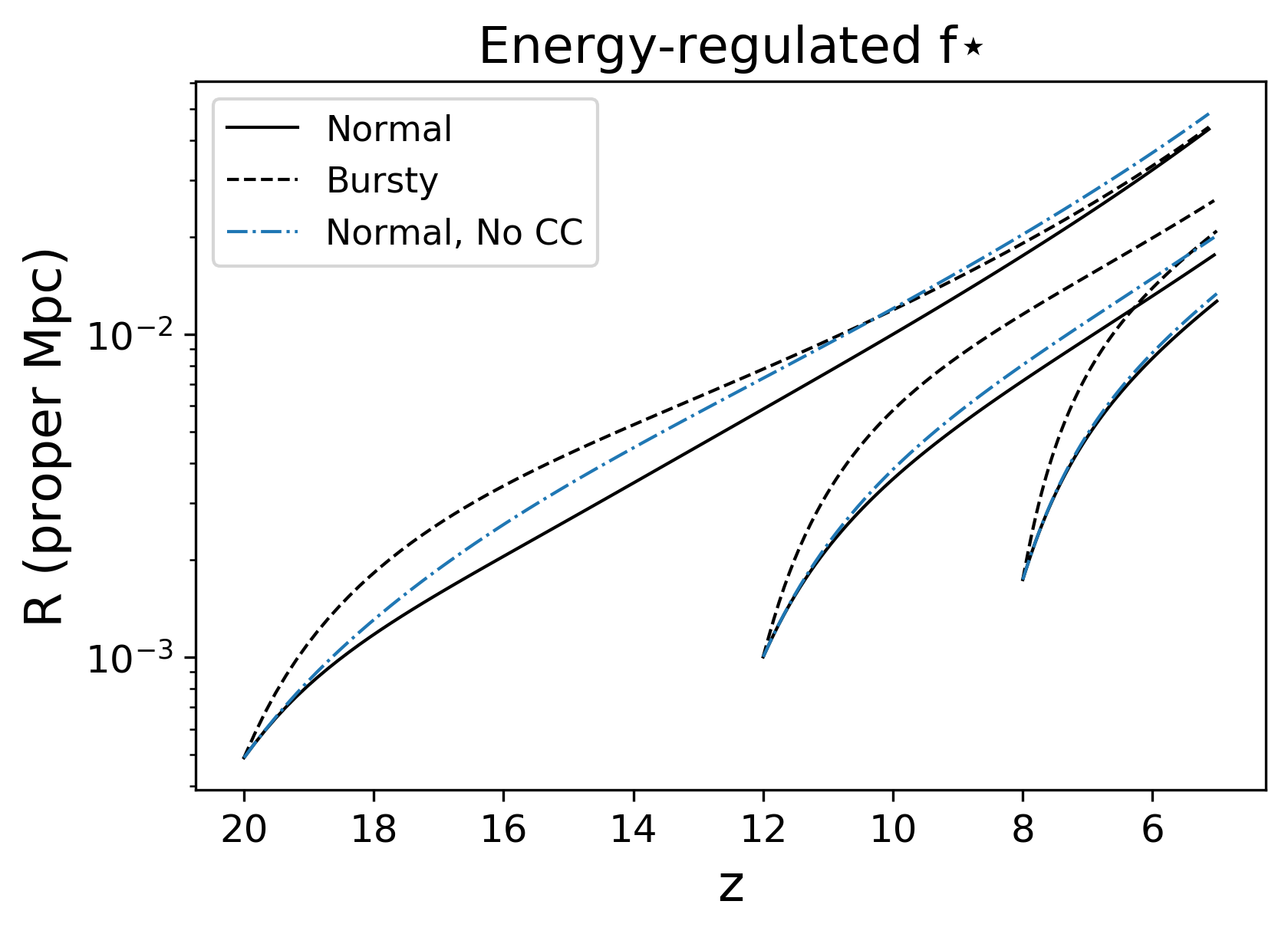}
    \includegraphics[width=\columnwidth]{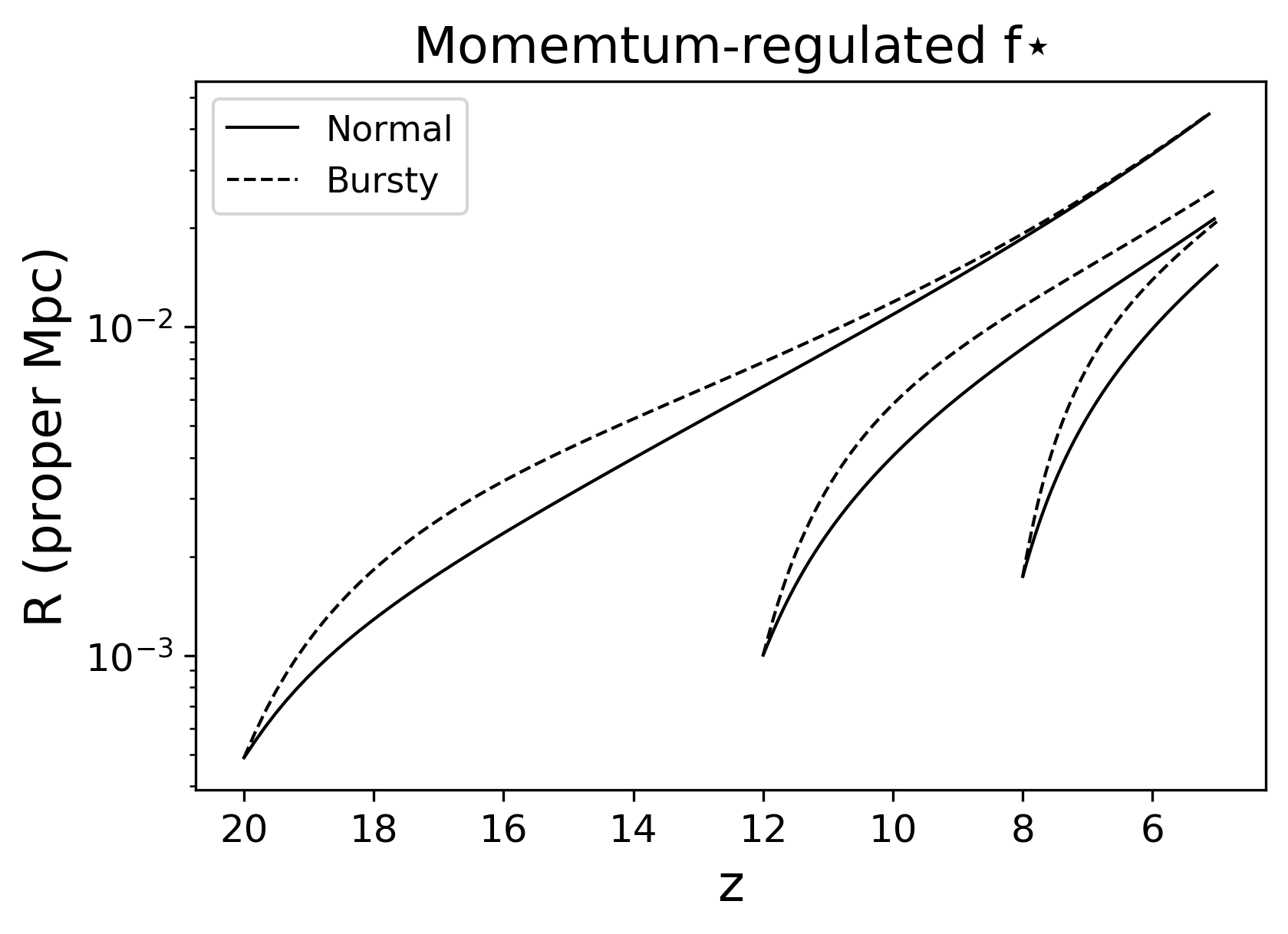}
    \includegraphics[width=\columnwidth]{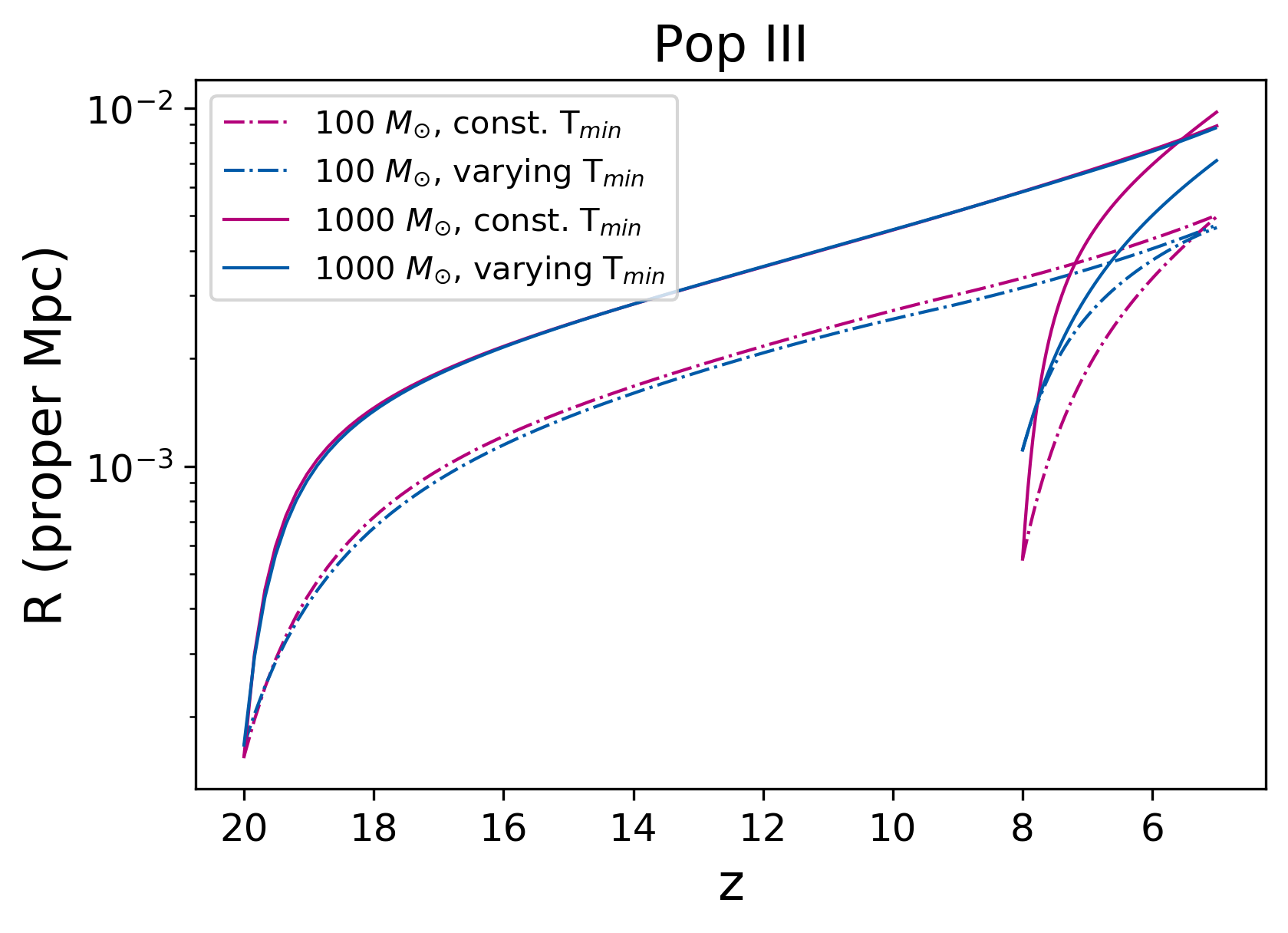}
    \caption{Evolution of the radius of an individual bubble starting at different initial redshifts for different SFE cases of the normal and bursty galaxy models: \emph{Top left:}  Constant $f_{\star}$ model. (0.1) \emph{Bottom left:} Momentum-regulated model. \emph{Top right:} Energy-regulated model (for this case, we also plot curves with energy losses from Compton cooling switched off). \emph{Bottom right:} Corresponding plots for several initial stellar masses for the Pop III model. We show the results for both a constant minimum virial temperature of 1000~K and a varying temperature which approximates the evolution of initial halo mass in a more complex history (see Section~\ref{ssec: burst}).}
    \label{fig:single_normal_bursty}
\end{figure*}

In comparison to \citet{tegmark_late_1993} and \citet{furlanetto_metal_2003}, we ignore a term that accounts for the uncertain fate of energy dissipated when particles are swept up. We find that, in the regime relevant to these galaxies, including this term
increases the wind radii by $\la 10$\%; ignoring it thus makes our filling factor estimates somewhat conservative. 

Furthermore, if the velocity of the wind slows down to the Universe’s expansion speed at $z=z_f$ and corresponding radius $R=R_f$, it will simply expand with the usual Hubble flow. If this condition occurs, we therefore set the comoving size to a constant. 

We initialize each wind at the redshift $z_{\rm init}$ the halo passes the relevant threshold to form stars. We assume that it instantly transforms a fraction $f_\star$ (evaluated at $z_{\rm init}$ and this threshold mass) into stars. We assume for simplicity that the wind begins at the halo's virial radius. This is reasonable because high-$z$ galaxies are compact, and the delay between SNe explosions and their winds reaching the virial radius is short ($\lesssim$ 10 \% of the Hubble time even if winds travelled at the circular velocity of the halo -- in our model the winds begin at much higher velocities.) We then distribute the energy released by the initial wave of supernovae equally between the hot interior (whose pressure drives the expansion) and the kinetic energy of the shell itself. Finally, we set the shell properties by demanding that it begins at the halo's escape velocity, unless the corresponding mass would exceed the halo's non-stellar baryonic mass. By running models across many choices for these parameters, we have verified that these assumptions do not affect the resulting radii of the winds by more than $\sim 25\%$.

\subsection{Example wind bubbles} \label{sec:examples}

Three panels in Figure \ref{fig:single_normal_bursty} show the evolution of the wind bubble radius $R$ for a single halo in the normal and bursty galaxy models starting at initial redshifts $z_{\rm{init}} = 8, 12,$ and $20$ for three SFE cases ($f_{\star} = 0.1$, energy-regulated and momentum-regulated). The parameters for the latter two models are summarized in Table \ref{tab:Table 1}.
In almost all cases, the bubbles initially expand rapidly before slowing, thanks to the sweeping up of ambient material. Energy loss from Compton cooling also contributes to the slowdown (and suppresses the final bubble radius), but it is not a large effect, as shown explicitly for the energy-regulated case in the upper right panel of Figure~\ref{fig:single_normal_bursty}. These haloes begin forming stars at masses of $\sim 10^{7-8} M_{\odot}$ and reach masses of $\sim (0.2,\,0.4,\,10)\times10^9 \ M_{\odot}$ at $z=5$ for $z_{\rm{init}} = 8, 12,$ and 20, respectively.

Overall, we find that winds that begin at earlier times grow larger than those launched at later times (even though they begin at smaller virial radii as $r_{\rm vir} \propto M_{h}[1+z]^{-1}$]). Although winds that begin at earlier times must propagate through higher-density material, their source halos grow steadily, so the halos forming at $z=20$ have much more mass (and stellar mass) than those at $z=8$. Moreover, even if these early winds ``stall," they continue to expand thanks to the Hubble flow (which is responsible for the rapid expansion seen at very late times in some $z = 20$ cases).

\begin{figure*}
    \centering
    \includegraphics[width=\columnwidth]{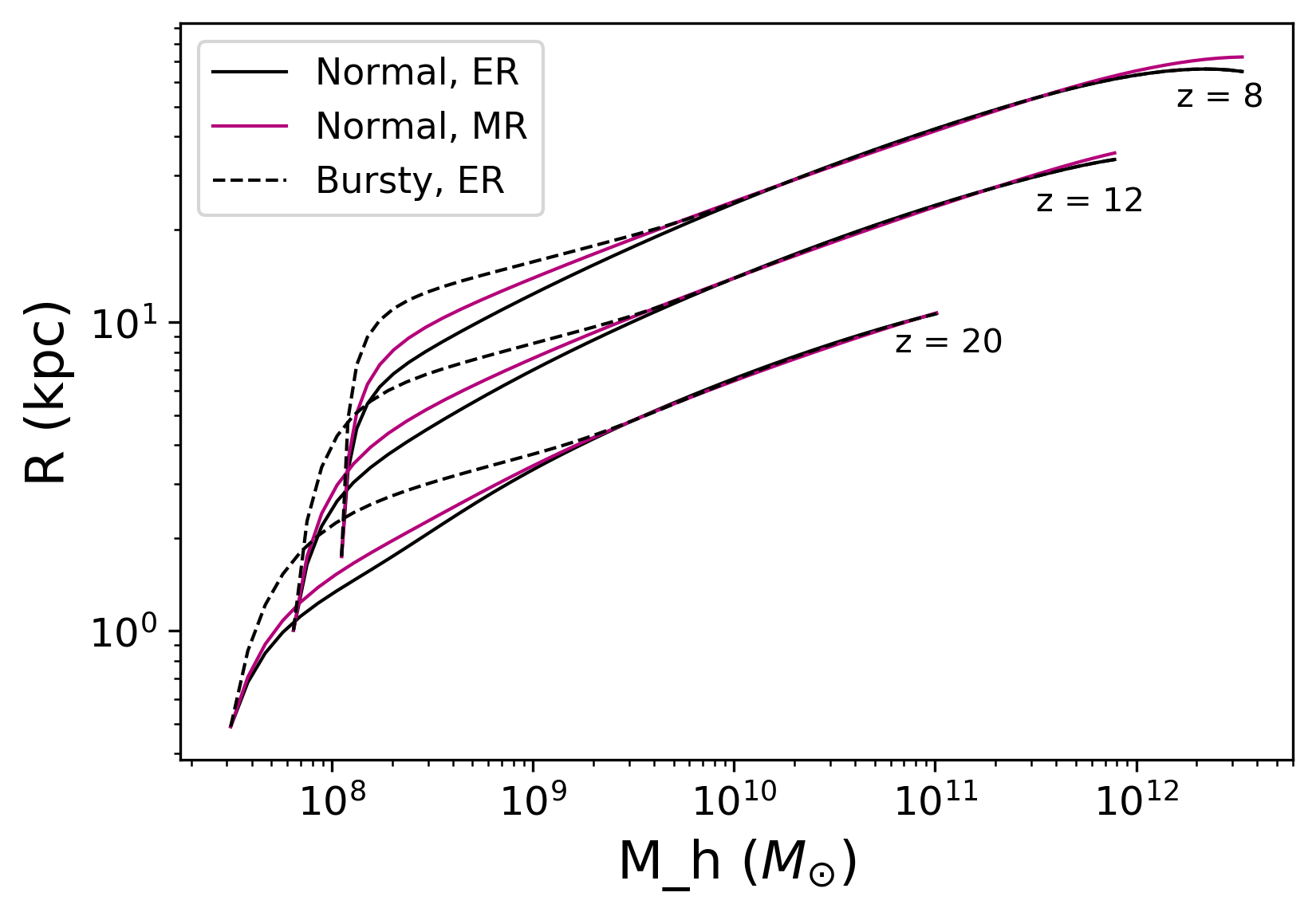}
    \includegraphics[width=\columnwidth]{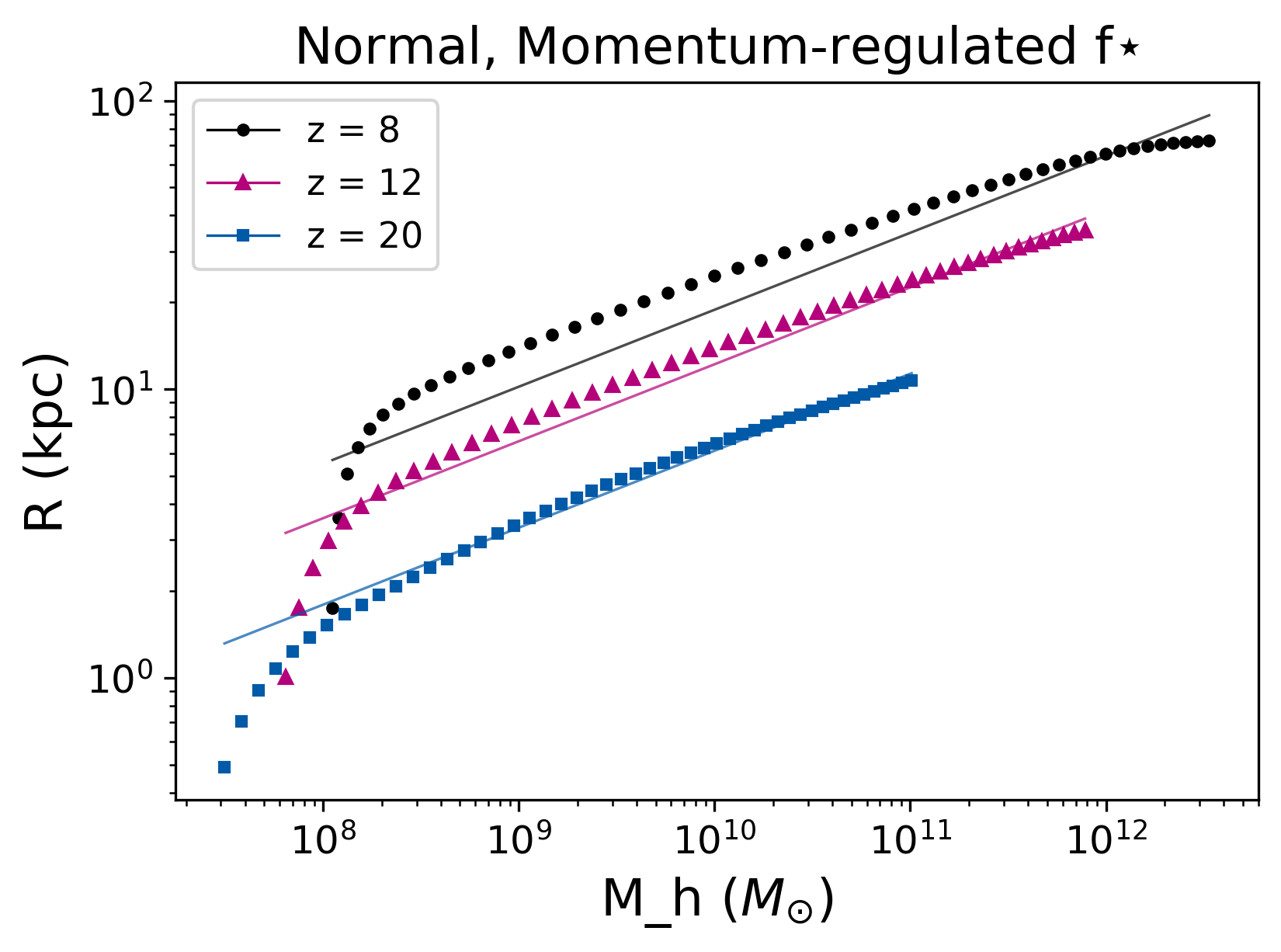}
    \caption{\emph{Left:} Radii of the wind bubble as a function of halo mass at various redshifts for the normal energy-regulated (ER) and momentum-regulated (MR) models as well as the bursty energy-regulated model. \emph{Right:} A power law fit of the form $R = A {M_h}^n$ to the normal momentum-regulated ($n = 4/15$) models at z = 8, 12, and 20.}
    \label{fig:finalradii}
\end{figure*}

It is worth noting that the initial expansion phase is much stronger for the bursty model, and for $z_{\rm init} = 8$ and 12, they reach significantly larger radii by $z = 6$ than the normal cases. This demonstrates the impact of increasing the star formation efficiency of the smallest halos, when the wind expansion begins (here $f_{\star,\rm min} = 0.03$). On the other hand, while the winds starting at $z_{\rm init} = 20$ are initially larger in the bursty models, they  approach the normal ones at later redshifts, because these halos have grown far above the low-mass limit in which the burstiness criterion matters. In addition, the difference is generally less significant for the momentum-regulated case because that model is less effective at suppressing star formation in small halos anyway, so the halos have a larger $f_\star$ throughout. 

Similar trends can be identified for the Pop~III model in the lower right panel of Figure \ref{fig:single_normal_bursty}. The key difference is that these wind bubbles are much smaller (note the different vertical axis in this panel). This is simply because they form many fewer stars, in contrast to the normal and bursty models, which continue to form stars throughout. Note as well that many of the Pop III model halos which form at later times actually grow larger than those that begin at earlier times -- opposite to the normal galaxies. This is because the surrounding medium has a higher density at early times, which makes it harder for the shell to expand. This effect is more significant for the Pop III model because all of the energy is injected at once. Thus, a wind that formed earlier gets stalled and those that form later, where the density is lower, quickly overtake them even though they have not had as much time to grow. This effect is less important for the normal galaxies because they continuously inject energy over a long period of time. 

The left panel of Figure \ref{fig:finalradii} plots the final radii of the wind bubbles as a function of source halo mass at three different redshifts. We show three different models at $z_{\rm{final}} = 8, 12,$ and $20$. These curves can be compared with Fig. 1. of \citet{furlanetto_metal_2003}. The qualitative agreement between our results and these older ones is reasonable (with one exception; see below). Note, however, that our models generally have smaller star formation efficiencies than assumed in that earlier work, especially in the normal energy-regulated model. The general trend of the radii rising sharply at low masses $\sim 10^8 M_{\odot}$ and then quickly slowing down to a relatively shallow slope is present in both figures. In addition, the radii increase significantly as the redshift decreases because the wind bubbles have had a longer time to expand. (In the newer models, the star formation efficiency also increases at lower redshifts.)

But there is one notable difference with earlier work -- our results do not show the rapid decline in radii at high halo masses present in \citeauthor{furlanetto_metal_2003}. This can likely be attributed to the simplifying assumptions  in our initial conditions: the more detailed, merger-tree construction of \citet{furlanetto_metal_2003} found that winds from high-mass halos would quickly be ``trapped" by the halo. This is generically expected because the halo binding energy increases more rapidly with halo mass than the energy supplied by supernovae. We do not attempt to model this regime in detail because our galaxy formation model is likely not accurate at high masses (because it ignores feedback from active galactic nuclei) and because halos at these high masses are sufficiently rare that winds from massive galaxies do not significantly affect our final results, which focus on the cumulative impact of the entire wind bubble population (see Section \ref{ssec:Q}).

Next we show that the qualitative behavior of the wind bubbles follows simple expectations. To that end, the right panel of Figure \ref{fig:finalradii} compares a power law of the form $R \propto M_h^n$, to our results. The power law index $n$ is set to $4/15$ here, which can be derived for the  momentum-regulated case by applying energy conservation at the asymptotic comoving radius of the wind. In detail, we assume that all the energy from supernova blastwaves goes into accelerating the swept-up IGM material to the Hubble flow velocity:
\begin{equation}
    E_{\rm SN} = \frac{1}{2} M_h \left(HR\right)^2 = \frac{1}{2}\frac{4\pi}{3} \rho\left(z\right) R^3 \left(HR\right)^2 \propto R^5.
    \label{eq:esn-fit}
\end{equation}
We then have $E_{\rm SN} \propto f_{\star} M_h$ (eq. \ref{eqn:Lw}) and $ f_{\star} \propto \left(M_h\right)^{\xi}$ (eq. \ref{eqn:eta}), so we expect $R \propto M_h^n$ where $n = \left(\xi + 1\right)/5$, which match the choices in Figure~\ref{fig:finalradii}. The estimate matches our momentum-regulated case relatively well, except at the low mass end, where the winds have only recently turned on so have not yet reached their maximum extent. It is clear from the success of this simple estimate that energy losses are not significant for the dynamics of these winds (though, again, we have not included the gravitational potential well at high masses accurately). Of course, the power law index will depend on the assumed star formation law, and it will become mass-dependent in our bursty model because $f_\star$ is not a pure power-law  in that case. (We also note that 
energy-regulated case is not as close a match to the expected index.)

\section{Results} \label{sec:results}

\subsection{Widespread enrichment of metal-carrying winds} \label{ssec:Q}

We next turn to our main result, the fraction of space filled by wind-driven metals. The volume filling factor of bubbles, $Q$, is
\begin{equation} \label{eqn:Q}
    Q = \int_{M_{\rm min}}^{\infty}dM_h\frac{dn}{dM_h}\frac{4\pi}{3}R_c^3,
\end{equation} 
with $R_c = R(1+z)$ obtained for each halo mass as in Figure \ref{fig:finalradii}. We emphasize again that our assumption of a roughly uniform distribution of metals with each wind bubble means that the volume filling factors calculated here serve as upper limits to the spread of metals.

\begin{figure}
    \centering
    \includegraphics[width=0.5\textwidth]{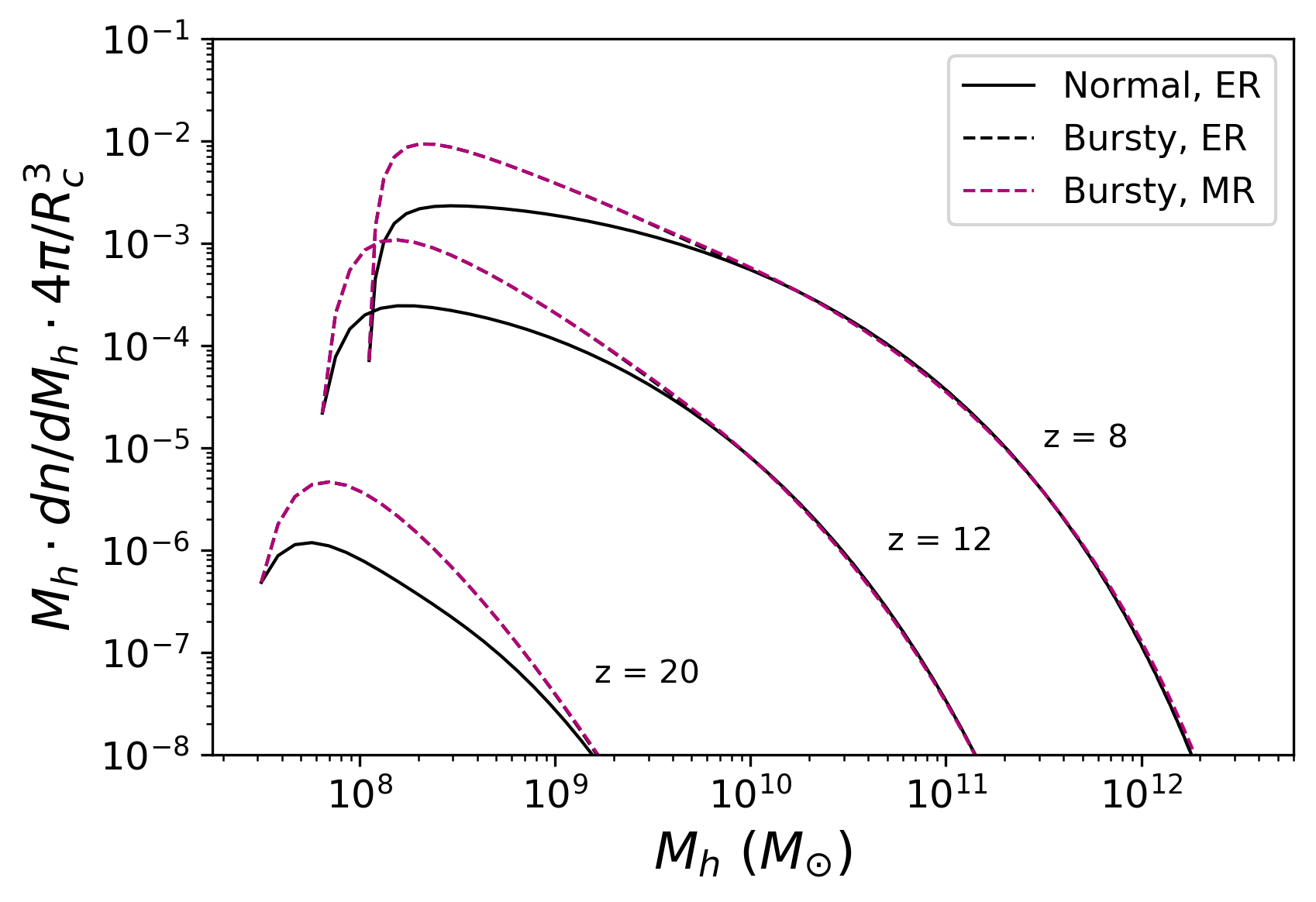}
    \caption{The integrand of equation~(\ref{eqn:Q}), or the contribution to the total volume filling factor from halos in each logarithmic mass range. The different line styles denote different star formation models: the minimalist energy-regulated model (solid curves), the bursty energy-regulated model (black dashed curves), and the bursty momentum-regulated model (magenta dashed curves); the two bursty cases are so close that they are hard to distinguish. In all cases, the integral is dominated by relatively low-mass halos.}
   \label{fig:Q_int}
\end{figure}

Figure~\ref{fig:Q_int} shows the integrand of this equation as a function of halo mass, thus providing the contribution of each mass range to $Q$. As mentioned briefly in Section~\ref{sec:examples}, we  see that halos with $M_h \lesssim 10^{11} M_{\odot}$ dominate over those at higher masses, especially at higher redshifts. Therefore, ignoring gravity in the highest mass halos does not significantly affect our results. We also see that the smallest halos are more important in the bursty models than in the normal models, because their star formation rate is strongly suppressed in our ``normal" energy-regulated model. 

\begin{figure*}
    \centering
    \includegraphics[width=\columnwidth]{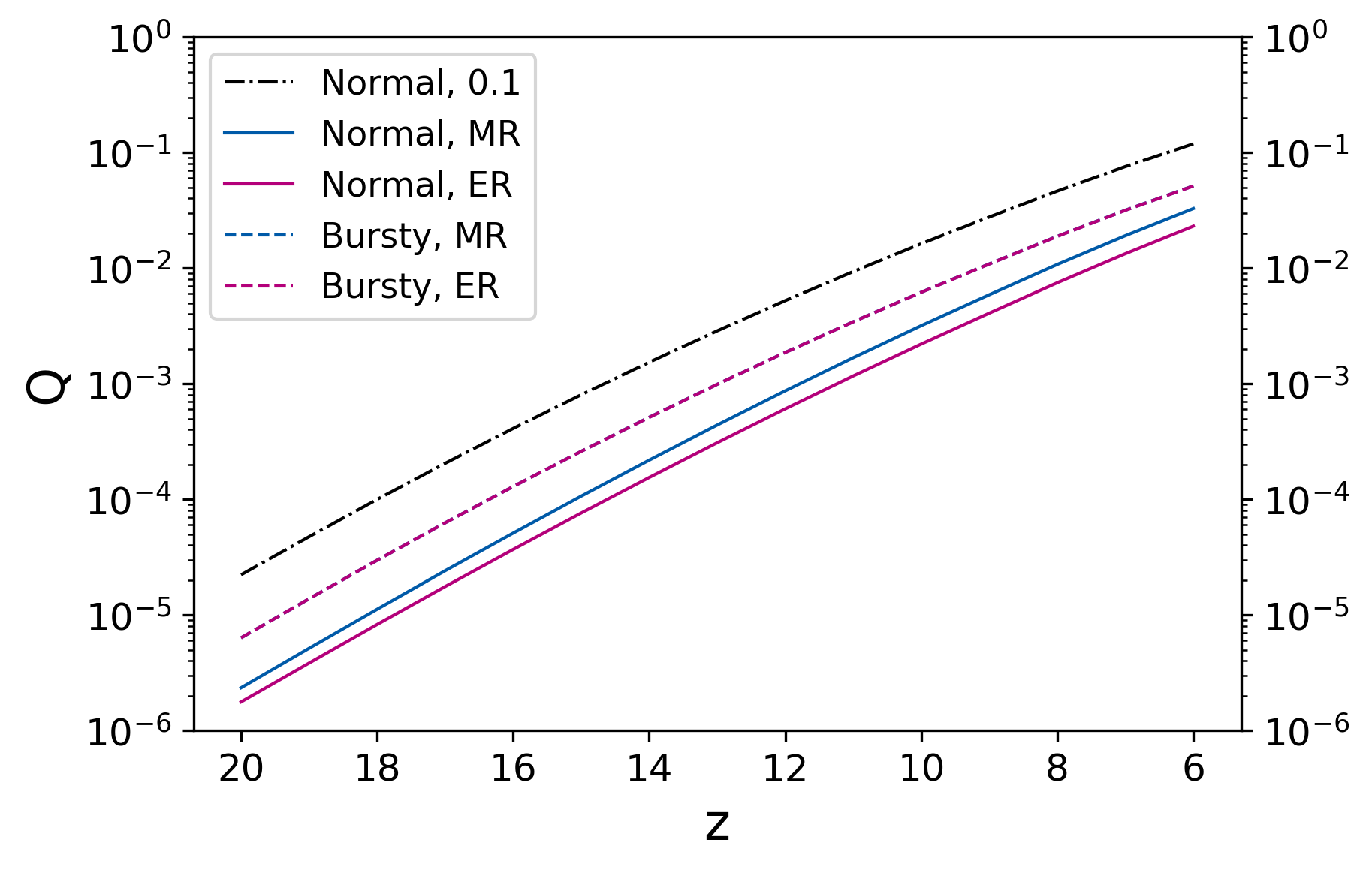}
    \includegraphics[width=\columnwidth]{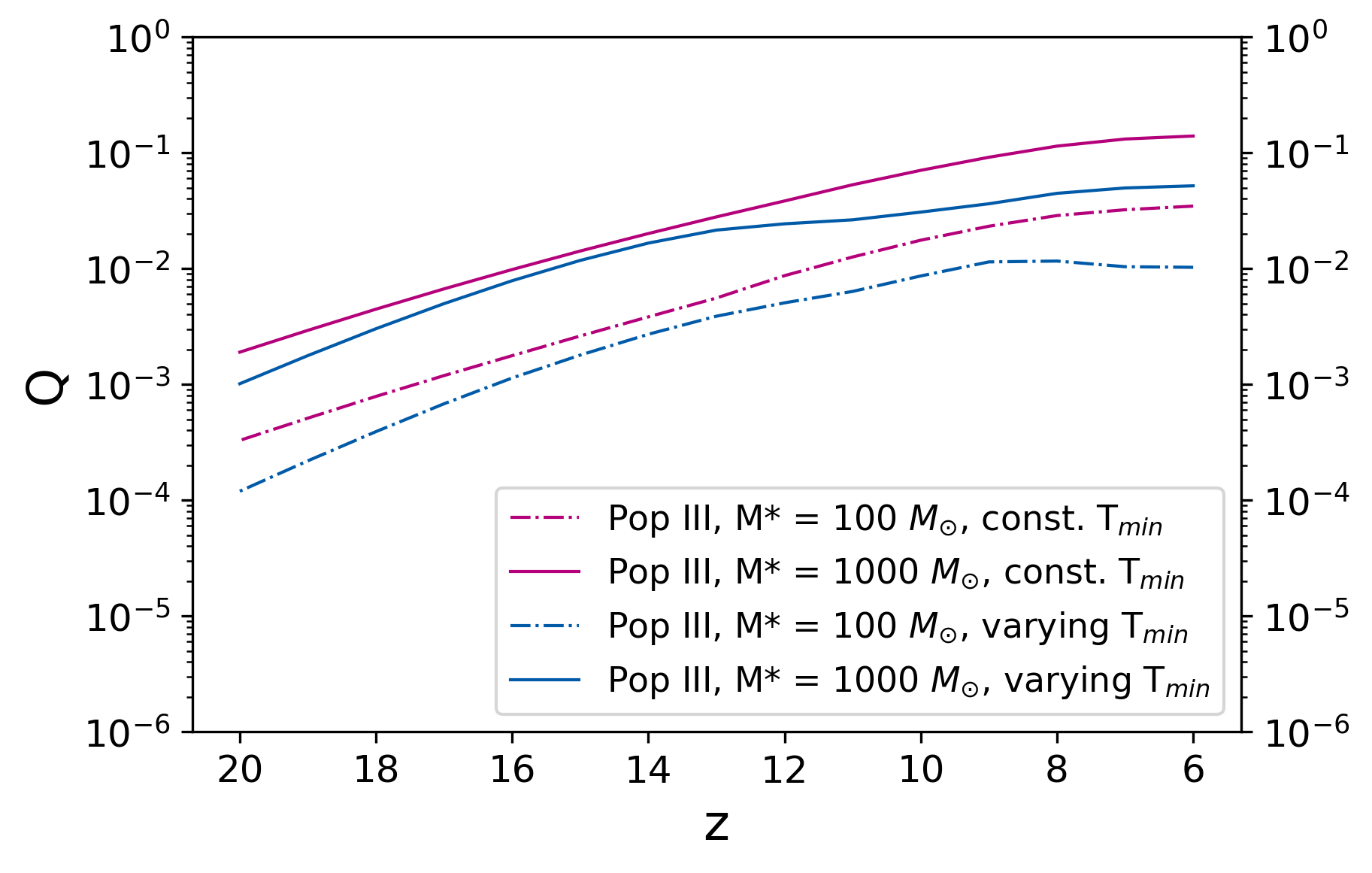}
    \caption{Evolution of the volume filling factor $Q$ in our models. \emph{Left:} ``Normal" galaxies, in which star formation proceeds over long timescales. From top to bottom, the solid lines take $f_\star=0.1$, momentum-regulated feedback, and energy-regulated feedback. The dashed curves impose a minimum $f_\star=0.03$, mimicking bursts in very small galaxies. \emph{Right:} Population~III models, in which star formation only occurs in a single burst. The curves assume different stellar masses in this burst and take two prescriptions for the mass at which this star formation occurs.}
    \label{fig:Q}
\end{figure*}

The left panel of Figure \ref{fig:Q} shows the evolution of $Q$ for several models of the star formation efficiency. From top to bottom, the solid curves take a constant $f_\star=0.1$, momentum-regulated feedback, and energy regulated feedback. The dashed curves show the bursty models, which closely overlap in the case of the momentum-regulated models. We remind the reader that the constant $f_\star=0.1$ model will not fit observed luminosity functions, but it is included for comparison with earlier results. We see that, regardless of the feedback prescription, $Q\sim0.1-1\%$ at $z\sim10$ and between $\sim1-10\%$ at $z\sim6$. 

The right panel of Figure \ref{fig:Q} shows the evolution of $Q$ for several Pop III models. Here, we show conservative and optimistic cases, corresponding to stellar masses of $100 \ M_\odot$ and $1000 \ M_\odot$, for both a constant threshold virial temperature and the fit to semi-analytic results. The values of $Q$ lie roughly between $\sim0.1$--$10\%$ (depending on $M*$) for all $z \lesssim 10$; by that time, relatively few new Pop~III stars are forming in these models. Note that the structure in the time-varying $T_{\rm min}$ curves comes from that threshold mass evolution. 

In both sets of models, metals can be \emph{relatively} widespread by $z=6$, but they are very unlikely to be pervasive, at least if they are spread by winds from star-forming galaxies. Only if the star formation efficiency of very small galaxies is increased dramatically can $Q \sim 1$ at $z=6$; even then it is difficult to imagine that it will be $\ga 10\%$ at $z \sim 10$. In particular, the power-law approximation in  Figure~\ref{fig:finalradii} suggests that each bubble's volume $\propto (\epsilon_K f_\star)^{3/5}$ at a fixed halo mass (which we have verified in our numerical models as well). To increase the filling factor to unity in our model, we would need to increase both of these factors to near unity as well.

Interestingly, we see that the overall shape of the curves for normal galaxies are quite similar, regardless of the particular star formation model. 
At all redshifts, $Q$ in the $f_\star=0.1$ model is about an order magnitude larger than that in the energy and momentum-regulated models. The bursty model, with enhanced star formation in small halos, does have significantly more widespread enrichment. This underlines the important role of the initial phases of star formation in early galaxies to metal enrichment and motivates further studies of it in greater detail.

Comparing the two panels of Figure~\ref{fig:Q}, we find that the Pop~III halos, even with the conservative $100 M_{\odot}$ estimate for the stellar mass,  contribute significantly to the total enriched volume, dominating at high redshifts. Although the stellar mass in each halo is very small in these cases, they have several advantages: they get very early starts (and so have time to reach their maximum size) and they form in abundant, low-mass halos. The filling factor from these sources is particularly large at early times, although the gap closes in at later redshifts because the continuous star formation in normal galaxies helps their bubbles continue to grow over time. However, we will see in the next section that the amount of enrichment in the Pop~III winds is very small. 

There are relatively few estimates of metal enrichment during this era, but we can compare our model to a few earlier calculations. 
\citet{furlanetto_metal_2003} used a more complex wind model but made similar assumptions to our $f_{\star}=0.1$ model (see their Fig.~5). We find  good agreement between the two calculations (when comparing to the \citet{furlanetto_metal_2003} calculation assuming a Scalo IMF and atomic cooling).
\citet{scannapieco_early_2002} also considered a very similar case (see their Fig.~1), finding $Q \sim 0.2$ at $z \sim 6$ (we note however that their results deviate from ours at high redshifts, showing "plateaus" in the filling factor for some models likely caused by the details of their more complicated cooling and wind prescriptions).
We conclude our simplifications to galaxy growth and the wind model are reasonable. However, we emphasize that a constant-$f_{\star}$ model is no longer viable in light of the observed luminosity function at $z \sim 6$--8, so the true filling factor is likely to be many times smaller than suggested by these earlier papers -- in the range $Q \sim 0.01$--0.05 rather than near unity. 

\citet{furlanetto_metal_2003} also considered enrichment from Pop~III halos, but they made far more optimistic assumptions about star formation in them, so it is difficult to compare directly, although the relative shapes of the curves are similar. However, \citet{jaacks_baseline_2018} used a simulation-based method to estimate the filling fraction. Their star formation model is much more sophisticated, but their results are roughly similar to our models with a stellar mass of $100\ M_\odot$ per halo, although our volume filling fraction is larger at late times. This may be because we assume that all the Pop~III wind bubbles remain independent; in reality, their source halos likely merge over time. In that case, we would overestimate their contribution at late times. 

\begin{figure*} 
    \centering
    \includegraphics[width=\columnwidth]{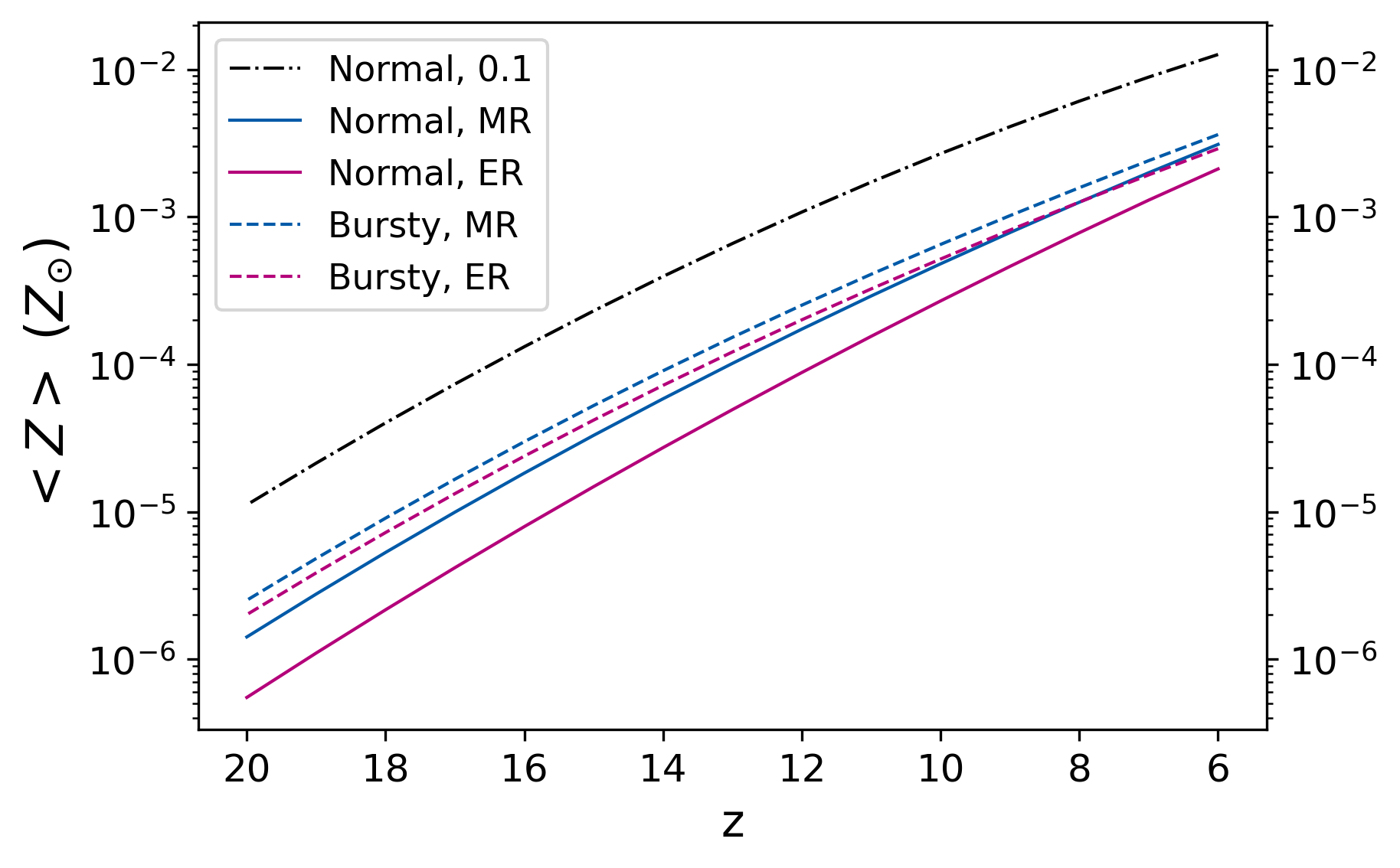}
    \includegraphics[width=\columnwidth]{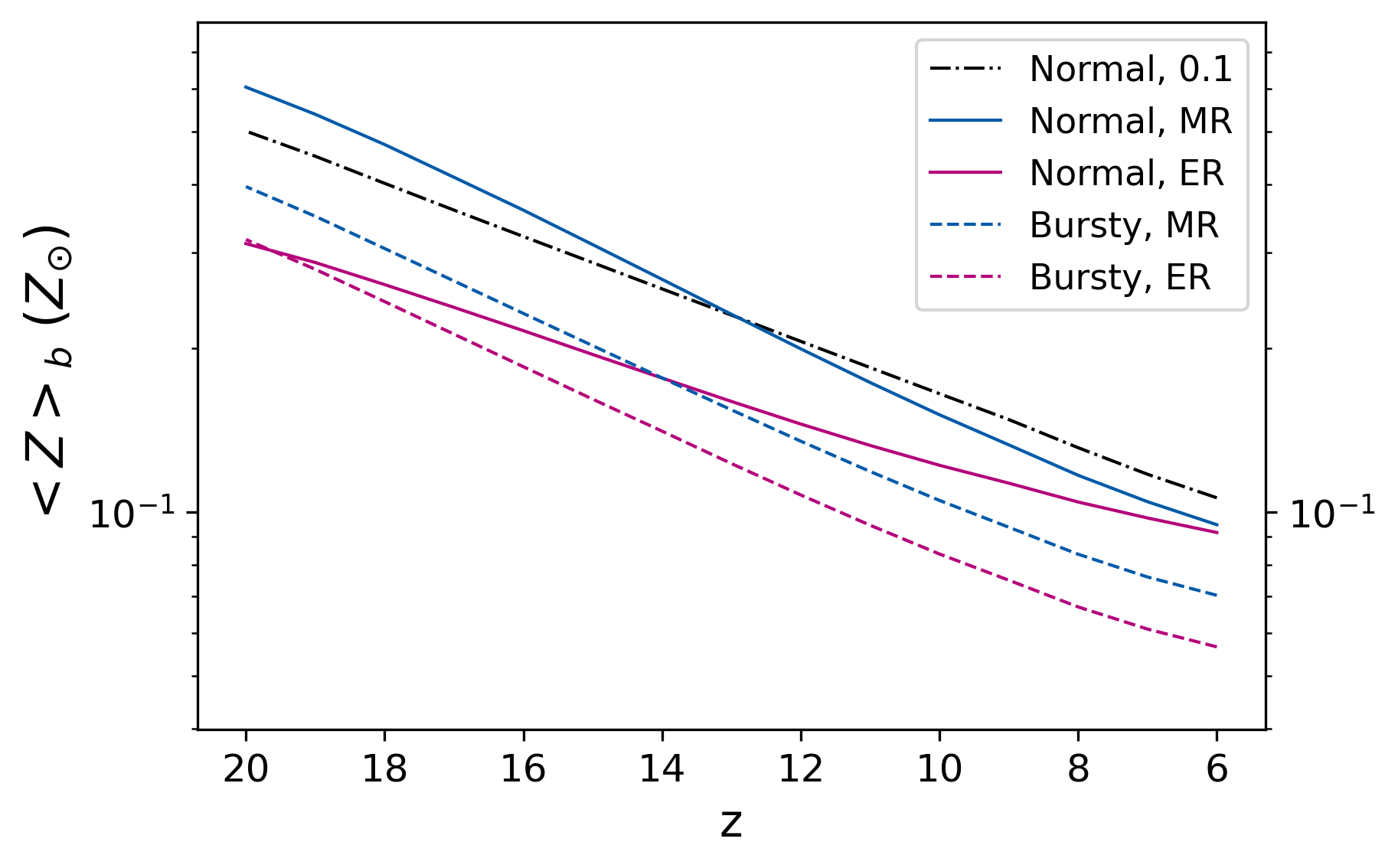}
    \caption{Evolution of the average metallicity of the Universe (\emph{left panel}) and of the wind-driven bubbles (\emph{right panel}) for several normal and bursty galaxy models, as in Fig.~\ref{fig:Q}. }
    \label{fig:Z}
\end{figure*}

\subsection{Average Metallicity} \label{ssec:Z}

Next, we consider how the average metallicity of the Universe $\left<Z\right>$, as well as that of the bubbles $\left<Z\right>_{b}$, evolve with time. The comoving mass density of metals produced is 
\begin{equation}
   \rho_{Z} = \int^{\infty}_{M_{\rm min}} M_{\rm metal}\left(M_h\right) \frac{dn}{dM_h} dM_h,
\end{equation}
where $M_{\rm metal}$ is the mass of a metal produced by a halo of mass $M_h$ (eq.~\ref{eqn:M_metal} with the fraction of stellar mass turned into metals, $y_Z = 0.03$ \citep{benson_galaxy_2010}, in place of $Y_i n_{\rm SN}$). The mean metallicity of the Universe is then $\left<Z\right> = \rho_{Z}/\bar{\rho}_{b0}$, and the mean metallicity of the enriched material inside bubbles is then $\left<Z\right>_{b} = \left<Z\right>/Q$. We report our results in units of solar metallicity, which we take to be $Z_{\odot} = 0.0196$ \citep{von_steiger_solar_2015} (Note that this recent estimate is larger than earlier measurements made by e.g. \citet{asplund_chemical_2009} and  \citet{lodders_abundances_2009}).

The left panel of Figure \ref{fig:Z} shows the evolution of $\left<Z\right>$ with redshift. As expected, we see that it increases with time as the total number of stars rises, reaching $\sim 1-4 \times 10^{-3}$ by $z = 6$ (excluding the constant-$f_{\star}$ case). This can be compared to Figure 10 in \citet{furlanetto_minimalist_2017}. The two figures are in good agreement, sharing similar shapes of the curves as well as very comparable $\left<Z\right>$ values. Furthermore, \citet{yates_cosmic_2021} finds $\Omega_{\rm metal} \sim 10^{-6}$ at these redshifts (their Fig.~1). Dividing this value by the baryon fraction and the solar metallicity, we get that $\left<Z\right> \sim 10^{-3}$, which is roughly consistent with the values we find here.

Meanwhile, we see from the right panel of Figure \ref{fig:Z} that the average metallicity inside the bubbles $\left<Z\right>_b$ falls with time. This is because, in most models, the wind bubbles expand fast enough that the metals are diluted more rapidly than new star formation produces them. This is natural for any wind. For example, the simple Sedov-Taylor blastwave has $R \propto E^{3/5}$. Because the input energy is proportional to the stellar mass -- and hence also the metal mass -- we would expect $\left< Z \right>_b \propto E/R^3 \propto M_*^{-4/5}$. Our winds are more complex than the simple Sedov-Taylor solution (with continuous energy injection and an expanding medium) but follow the same qualitative trend. 

We also note that the bubble metallicity is rather large. This is at least partly a consequence of our assumption that \emph{all} metals are ejected from their source galaxy; in reality, a large fraction will likely cycle back into the interstellar medium without fully escaping.

For the Pop~III models, the metallicities are typically much smaller, with $\left<Z\right>$ ranging between $\sim 10^{-7} - 10^{-8}$: these winds are efficient at spreading metals around the Universe, but only at extremely low levels. This would remain true even if Pop~III supernovae had larger metal yields. 

\subsection{Metal-line absorption systems} \label{ssec: dndx}

At present, the only way to observe metal enrichment in the early Universe is through metal line absorption systems seen against luminous background quasars. This does not directly probe the filling factor, but rather the incidence of (strong) enrichment along each line of sight. To compare to these observations, in this section we therefore estimate the incidence of such absorbers in our models. 

\begin{figure}
    \centering
    \includegraphics[width=0.5\textwidth]{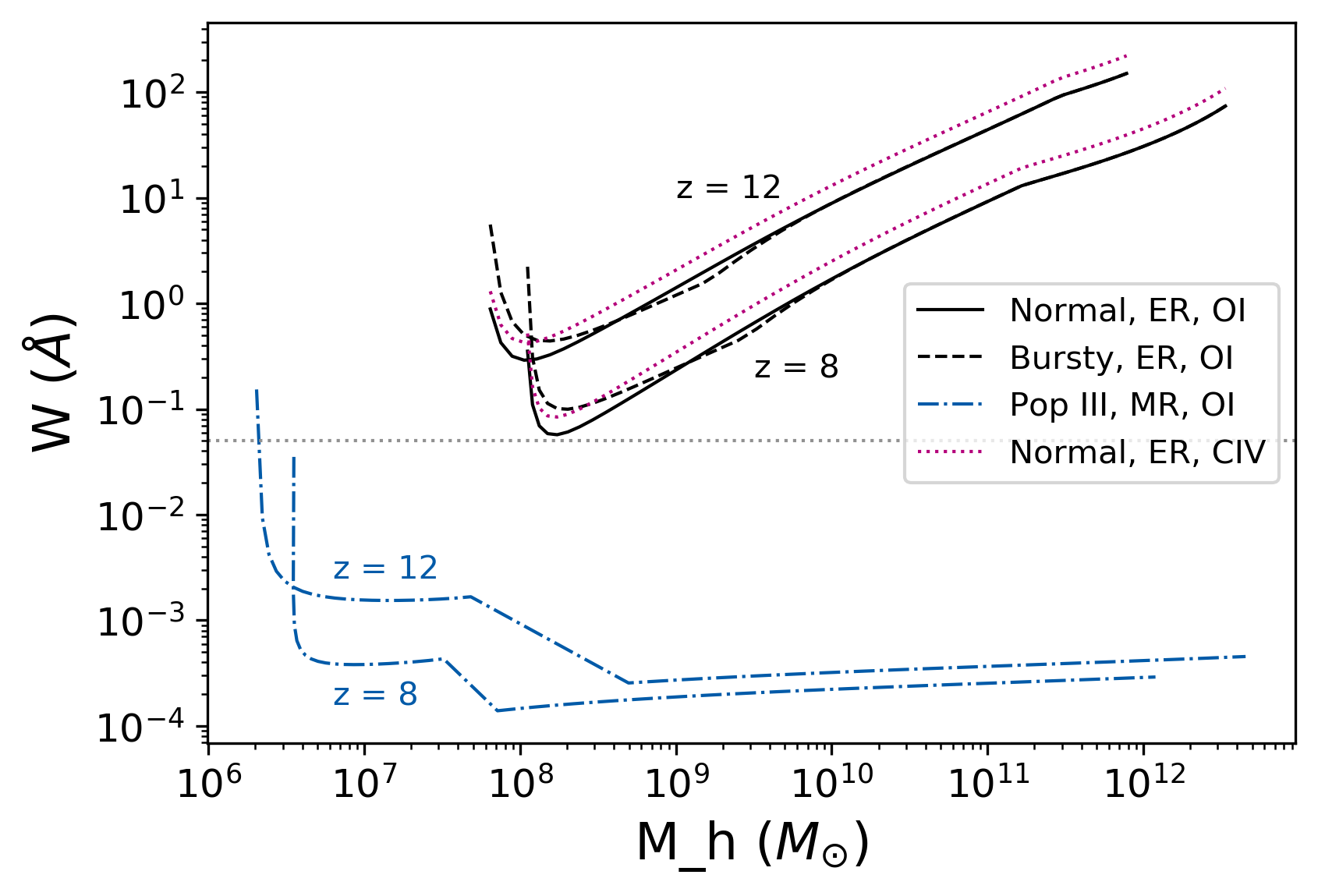}
    \caption{Equivalent width of the OI line produced by wind bubbles as a function of halo masses at several redshifts and for different galaxy models. The dotted line at 0.05\AA \ represents the sensitivity limit of \citet{becker_evolution_2019}. The other dotted lines show the equivalent width of the CIV line instead.}
    \label{fig:EW}
\end{figure}

To do so, we must specify how metals are distributed within the wind bubble. We assume for simplicity that every line of sight through a wind bubble has a metal absorber and that all have equal column density. In reality, there may be a non-uniform distribution of metals, which would generally lead to fewer (but stronger) lines. Furthermore, we assume that all metals are in the ionization state of the observed line. Therefore, our model should only be taken as an upper limit to the true incidence of absorbers. 

\begin{figure*}
    \centering
    \includegraphics[width=\columnwidth]{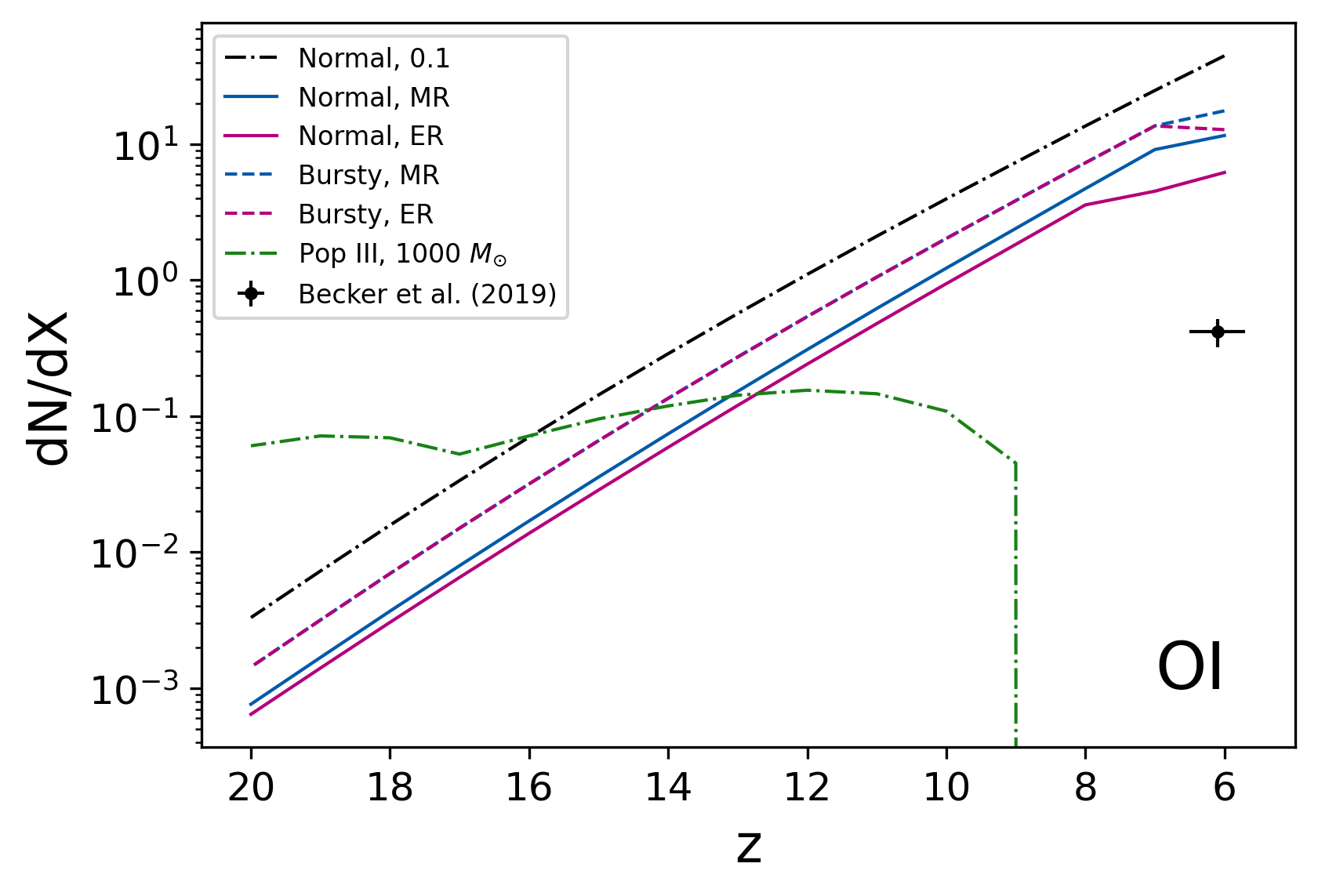}
    \includegraphics[width=\columnwidth]{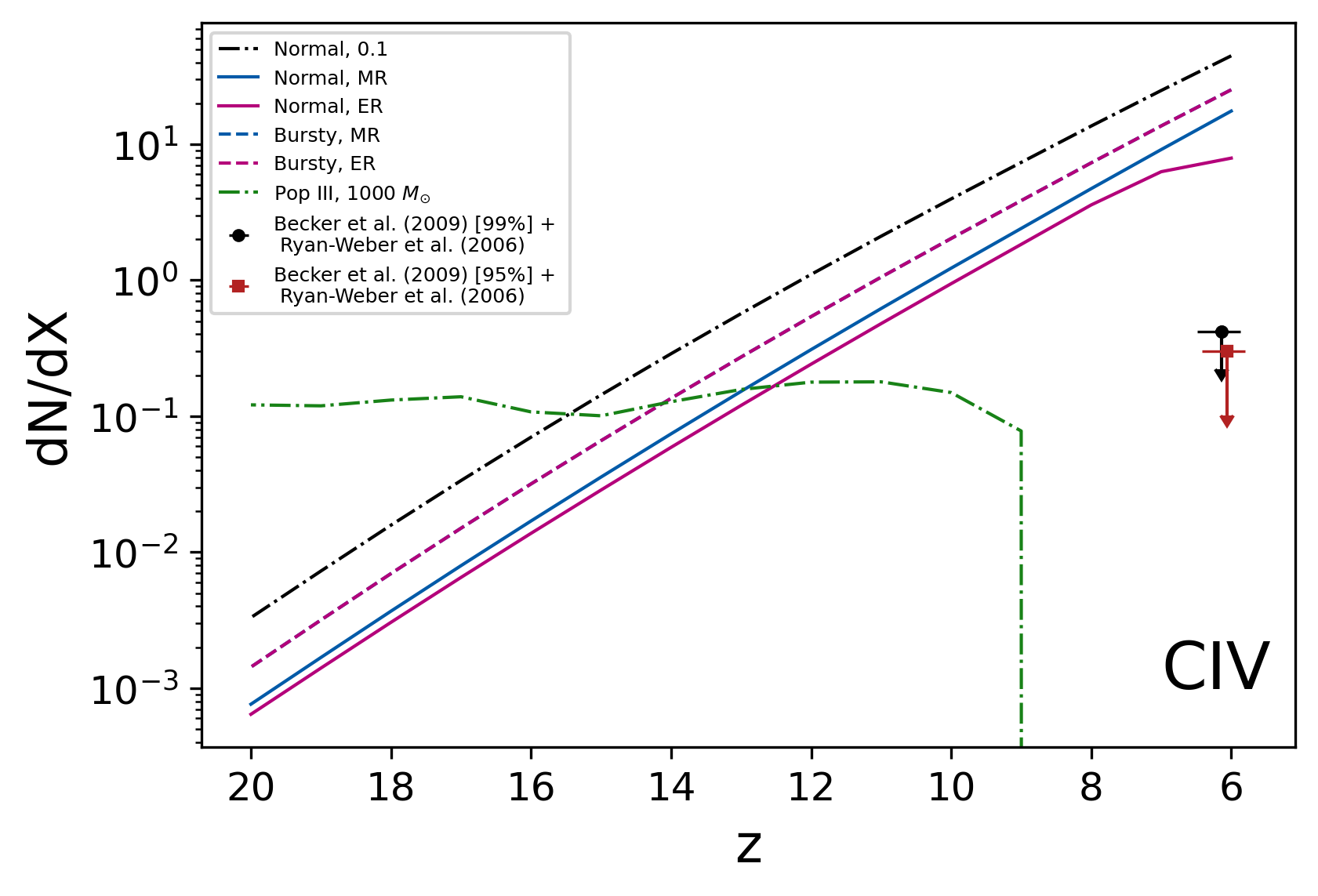}
    \caption{Comparison of predicted line incidence for OI (\emph{left panel}) and CIV (\emph{right panel}) to observations at $z\sim 6$ . In each case, the theoretical curves include only those systems that exceed the limiting equivalent width of the observations. We show several different model predictions. For OI, the data point is from Table 2 in \citet{becker_evolution_2019}, for $z = 5.7-6.5$.  For CIV, the data points are a combination of the upper limits of the number of absorbers (95\% and 99\% confidence intervals) from four sightlines in \citet{becker_high-redshift_2009} and one absorber detected in \citet{ryan-weber_intergalactic_2006}. The Pop III model shown here is for constant $T_{\rm min}$.}
   \label{fig:dNdX}
\end{figure*}

We can then use the metal yields from Section \ref{ssec:MP} to estimate the equivalent width of each wind bubble, shown in Figure \ref{fig:EW} for the OI line. We show the equivalent width as a function of halo mass for several galaxy models at $z = 6$, $12$, and $20$.  We see that $W$ is relatively insensitive to the star formation law in the normal and bursty galaxy models: equation~(\ref{eqn:EW}) shows that $W \propto X_\star M_h/R^2 \propto 1/(\xi + 1)$ using the fits in equation~(\ref{eq:esn-fit}), which does not depend strongly on the star formation law (we emphasize again that given the numerous simplifications made in our models, the exact values of $W$ should only be taken as estimates.). The Pop~III models are all much lower, however, because they have much smaller stellar masses driving them (which are also independent of the halo mass). 

The horizontal dotted line in Figure~\ref{fig:EW} at 0.05~\AA \ corresponds to the observational cutoff below which the absorption cannot be detected in \citet{becker_evolution_2019}. We see that nearly all the wind bubbles around ``normal'' galaxies are likely detectable, except in the Pop~III models.

Under these simplifying assumptions, the number density of absorption  lines per unit path length, $dN/dX$, is an integral over the cross-sectional area of the bubbles:
\begin{equation} 
    \frac{dN}{dX} = 90 \ {\rm Mpc} \  \left(\frac{7}{1+z}\right)^2\int dM_h\frac{dn}{dM_h}\pi R_c^2. 
\end{equation}
The prefactor  comes from $X$, the absorption path length interval, originally defined in equation~(4) of \citet{bahcall_statistical_1969}, where
\begin{equation}
    \frac{dX}{dz} = \left(1+z\right)^2\frac{H_0}{H(z)}.
\end{equation}
We set the limits of the integral to include only those halos whose wind bubbles exceed the threshold for line detection (e.g., the horizontal line in Fig.~\ref{fig:EW}). 

The evolution with redshift of $dN/dX$ for OI for several different galaxy models is shown in the left panel of Figure \ref{fig:dNdX}. For the normal galaxy population, the incidence of absorbers rises rapidly with redshift, so that their relative abundance more or less follows that of $Q$ in Figure \ref{fig:Q}. However, the Pop III model -- even taking an extreme case in which each halo forms $1000 \ M_{\odot}$ of stars -- remains relatively constant and sharply falls off at $z \sim 10$. This follows from the discussion of $W$ -- although the volume occupied by the wind bubbles, and thus $Q$, from these galaxies are significant, the metal yields are low and thus only a small fraction of the lowest mass halos have detectable absorption. By $z \sim 9$, they completely fall below the observational limit. The right panel of Figure~\ref{fig:dNdX} shows analogous results for CIV; given our extremely simplified assumptions about the metal distribution, the curves are very similar, with the only differences due to the different atomic yields and line properties. (Recall that we assume these species dominate the ionized fractions, which is not realistic!)

The extent of metal enrichment can be studied through quasar spectra and recent observations have pushed back the redshift limits to $z \sim 5$--$6$. As one example, \citet{becker_evolution_2019} presents a survey of 74 systems between $z = 3.2 - 6.5$ with O I $\lambda1302$ absorption with equivalent width $W > 0.05$ \AA.  \citeauthor{becker_evolution_2019} present constraints on $dN/dX$ at $z = 5.7-6.5$, with $dN/dX = 0.421$, shown  on the left panel of Figure \ref{fig:EW}. On the right panel, we compare to the combined detections of CIV absorbers along sightlines to quasars in \citet{becker_high-redshift_2009} and \citet{ryan-weber_intergalactic_2006}. 

We see that all of our normal galaxy models fill a sufficient volume to explain the current observations of OI and CIV line number densities at $z \sim 5-6$. Typically we overproduce the lines by an order of magnitude or so, which is not surprising in light of our very optimistic assumptions of a uniform metal distribution and that all atoms are in the appropriate ionic state. Nevertheless, the observations are below the upper limits predicted by the simplified models presented here, and thus they do not contradict the current theory. However, future data and better models of the connection between wind bubbles and absorbers are needed to make any stronger conclusions. 

We also find that absorbers from the Pop III model are rare except at the highest redshifts. Thus, although they may contribute to a low-level enriched background, they cannot explain the observed systems at $z \sim 5-6$. This would remain true even if we treated them as more energetic pair instability supernovae: the metal yields can be larger, but the increased explosion energies dilute them significantly.

\subsection{Effects on the CMB} \label{ssec:CMB_effects}

While metal lines surveys provide one probe of the extent of metal enrichment, they are challenging to perform and can only probe a small number of lines of sight at the highest redshift. It is therefore worth considering alternative observational constraints on high-$z$ winds. In this subsection,  we consider the Sunyaev-Zeldovich (SZ) effect as a possible alternative approach to detect the effects of galactic winds. This occurs when CMB photons scatter off of hot electrons along the line of sight, imprinting a fluctuation (with a characteristic spectral distortion) on the CMB \citep{rashid_alievich_sunyaev_small-scale_1970}. It has proven to be a powerful tool in observational cosmology, especially for searching for galaxy clusters (e.g., \citealt{planck_collaboration_planck_2014, bleem_galaxy_2015}). But any source of hot electrons can imprint SZ fluctuations, including high-redshift star formation \citep{oh_sunyaev-zeldovich_2003} and winds from quasars and galaxies (e.g, \citealt{majumdar_sunyaev--zeldovich_2001, platania_sunyaev-zeldovich_2002, white_simulating_2002}).

In this section, we estimate the amount of CMB heating due to the Compton cooling of the hot gas inside the wind bubbles. Following \citet{oh_sunyaev-zeldovich_2003}, we focus on estimating the Compton-$y$ parameter, which characterizes the total average spectral distortion along a line of sight, to look at how the magnitude of the SZ effect depends on some of our input parameters. In our model, the energy density injected by supernovae is
\begin{equation}
    \epsilon_{\rm SN} = \omega_{\rm SN}f_{\rm comp} \tilde{f}_{\star}f_{\rm coll}\Omega_b {\rho}_{\rm crit},
\end{equation}
where $f_{\rm coll} = \rho_{\rm halo}/\bar{\rho}_m$ is the fraction of matter contained in star-forming halos, $\tilde{f}_{\star}$ is the mass-averaged star formation efficiency, and a fraction $f_{\rm comp}$ of the energy is injected into the CMB. Thus, the energy input per baryon can be written
\begin{equation}
    \frac{\epsilon_{\rm SN}}{\bar{n}_b} 
    \approx0.25 \ { \rm eV} \left(\frac{\omega_{\rm SN}}{10^{49} {\rm erg}/M_{\odot}}\right)\left(\frac{f_{\rm comp}}{0.1}\right)\left(\frac{f_{\star}}{0.01}\right)\left(\frac{f_{\rm coll}}{0.05}\right)
\end{equation}
where $\bar{n}_b = \Omega_b \rho_{\rm crit} / m_p$ is the average number density of baryons. 

As a simple estimate of the amplitude of the SZ effect, we compute the Componization parameter, or the average $y$-distortion \citep{as_kompaneets_establishment_1957}, which is the dimensionless timescale for the collision of a photon in an electron field. It equals the integrated electron pressure along a line of sight and can be approximated as the ratio of energy density injected through Compton cooling of the wind bubbles to the energy density of the CMB \citep{zeldovich_interaction_1969}, with $y = -1/2*\Delta T / T$. We find  (see Section 3.3 of \citealt{oh_sunyaev-zeldovich_2003})
\begin{eqnarray}
    y & \approx &  10^{-8}\left(\frac{7}{1+z}\right)\left(\frac{f_{\rm comp}}{0.1}\right) \left(\frac{f_{\star}}{0.01}\right)  \nonumber
    \\ & & \times \left(\frac{f_{\rm coll}}{0.05}\right)\left(\frac{\omega_{\rm SN}}{10^{49} \rm erg/M_{\odot}}\right)
    \label{eqn:y-est}
\end{eqnarray}
Note that the amplitude is proportional to $f_{\star}$ as well as to $f_{\rm comp}$, the fraction of SN energy which is lost through Compton cooling. The latter can be approximated as $u_{\rm comp}/u_{\rm SN}$ where $u_{\rm SN}$ is the total energy density injected into the winds by supernovae, and $u_{\rm comp}$ is the energy density transferred from the winds to the CMB.  

Equation~(\ref{eqn:y-est}) suggests that the CMB distortion due to high-$z$ winds is quite small. The COBE FIRAS instrument constrained $y < 1.5 \times 10^{-5}$ \citep{fixsen_cosmic_1996}, while \citet{khatri15} reduced the limit to $\la 2 \times 10^{-6}$.  Our result is also well below the predictions of \citet{oh_sunyaev-zeldovich_2003}, because they made much  more optimistic assumptions about the efficiency of Pop~III star formation. 

For a more detailed estimate of the distortion, we trace the energy lost by our wind bubbles to the CMB, $E_{\rm comp}$, in a process similar to that used to obtain Figure \ref{fig:finalradii}. The resulting energy density is
\begin{equation}
    u_{\rm comp} = \int_{M_{\rm min}}^{\infty}dM_h\frac{dn}{dM_h}E_{\rm comp}\left(M_h\right).
\end{equation}
This expression allows us to obtain the redshift evolution of the Compton cooling rate from the wind bubbles for all galaxy models. The total energy injected increases with cosmic time, because bubbles continue to grow (and lose energy) as galaxies form more stars. 

The total energy density $u$ of the CMB evolves as
\begin{align}
    \frac{du}{dz} = - 4 H\left(z\right)\frac{dt}{dz}u\left(z\right) + \frac{du_{\rm comp}}{dz} \left(1+z\right)^3.
\end{align}
The first term simply represents the decreasing density due to the expanding Universe and the second term accounts for the energy injected from the wind bubbles (in proper units). Using $u(z) = aT(z)^4$, the temperature distortion $\Delta T$ caused by the inclusion of the second term can be calculated. 

\begin{figure}
    \centering
    \includegraphics[width=0.5\textwidth]{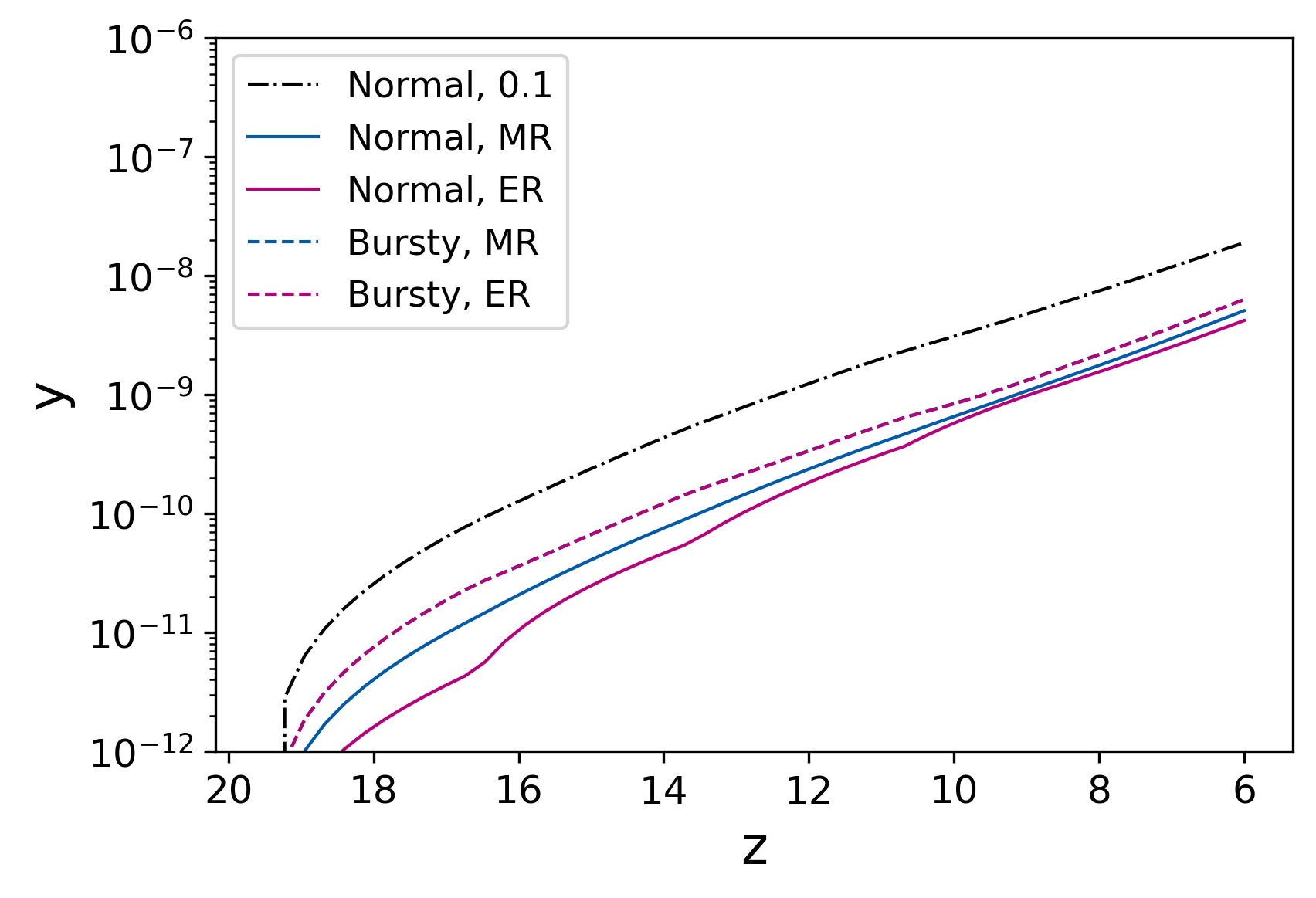}
    \caption{Evolution of the Compton-$y$ parameter for several of our galaxy models.}
    \label{fig:y}
\end{figure}

Figure \ref{fig:y} shows the resulting $y$-distortion to the CMB. The Compton $y$-parameter increases from  $\la 10^{-11}$ to $\sim 10^{-8}$ in our models that are calibrated to the luminosity function, as expected from our simple estimate. 
The $y$-distortions produced by the Pop III models are even smaller, ranging from $10^{-13}$ to $10^{-10}$, although these could increase by at least an order of magnitude if the Pop~III supernovae are particularly energetic (as with pair instability supernovae).

The smallness of the induced $y$-distortion will make it difficult to measure on average, because it cannot be separated from other sources (like the hot IGM at lower redshifts). \citet{oh_sunyaev-zeldovich_2003} pointed out, however, that the distortion will not be homogeneous, thanks to the clustering of the wind sources: a line of sight passing through an overdense area of high-$z$ galaxies will observe a large distortion. This is captured in the SZ power spectrum $C_{l}\left(y\right)$ (see eq.~13 of \citealt{oh_sunyaev-zeldovich_2003}). The angular structure should be similar to those in \citet{oh_sunyaev-zeldovich_2003}, as our source models are not dramatically different. Unfortunately, the clustering signal is proportional to $y^2$, which is  small in our models. 

\section{Discussion}  \label{sec:discussion}

In this paper, we studied the extent of metal enrichment of the intergalactic medium at $z \ga 6$ from winds driven by star-forming galaxies. We combined a simple model of star formation that matches observations at $6 \la z \la 8$ \citep{furlanetto_minimalist_2017} with a simple model of wind expansion \citep{tegmark_late_1993, furlanetto_metal_2003} that allowed us to study several star formation scenarios. 

We find that, when the galaxy model is calibrated to existing observations of high-$z$ galaxies, the volume filling factor $Q$, representing the fraction of the Universe enriched by metals through winds, only reaches at most $\sim 1 - 10 \%$ at $z \sim 6-8$. This is true even assuming relatively efficient star formation in small halos. Therefore, although it is easy to imagine galactic winds permeating a substantial fraction of space, it is challenging to get a majority of the Universe enriched. This implies that the process of chemical enrichment is highly nonuniform, so that one can imagine metal-free star formation taking place at late times, at least in principle. Our predictions for the filling factor are somewhat smaller than many earlier works (e.g., \citealt{furlanetto_metal_2003}), because those works made optimistic assumptions about the overall star formation efficiency of high-$z$ galaxies (and so do not reproduce the observed luminosity function.

Despite this relatively low level of enrichment, we find that normal galaxies can easily account for the few observations of $dN/dX$ currently available at $z \sim 5-6$ \citep{becker_high-redshift_2009, becker_evolution_2019, ryan-weber_intergalactic_2006}, although we did not attempt to model the distribution of metals or ionic species within each wind bubble. Nevertheless, the inhomogeneity implies that metal lines cannot be used ``out of the box'' to study reionization but will require simultaneous modeling of the metal and ionization distributions \citep{oh_probing_2002, hennawi_probing_2021}. 

In addition to the ``normal'' galaxy population, we also considered enrichment from Pop~III stars forming in minihalos. While each such halo only forms a few stars, we find that they are sufficiently numerous to provide a comparable filling factor to normal galaxies at later times, with $Q \sim 1\%$ in reasonable models. At early times ($z \ga 10$), enrichment from these sources can even dominate over normal galaxies, because they are so widespread. However, the star formation driving these episodes is at a very low level, and the resulting enriched regions have a very low average metallicity, well below the limits of metal-line systems. 

Finally, we considered the Sunyaev-Zel'dovich distortions produced by Compton cooling of the hot wind bubbles by calculating the $y$ distortion. We found that, for all models, the $y$ values are significantly smaller than the upper limit from COBE FIRAS \citep{fixsen_cosmic_1996} as well as those predicted by \citet{oh_sunyaev-zeldovich_2003}, because we made much less optimistic assumptions about the efficiency of early star formation. Even if this signal could be detected, it would require separation from low-$z$ sources of SZ distortions. 

\section*{Data Availability}

No new data were obtained as part of this work. Results used to generate the figures are available from the authors upon reasonable request.

\section*{Acknowledgments}

We thank G.~Sun and the anonymous referee for comments that improved this manuscript. This work was supported by the National Science Foundation through award AST-1812458. In addition, this work was directly supported by the NASA Solar System Exploration Research Virtual Institute cooperative agreement number 80ARC017M0006. We also acknowledge a NASA contract supporting the ``WFIRST Extragalactic Potential Observations (EXPO) Science Investigation Team" (15-WFIRST15-0004), administered by GSFC. N.~Y. thanks the UCLA Department of Physics \& Astronomy for support during its 2021 Undergraduate Summer Research Program.

\bibliographystyle{mnras}

\end{document}